\documentclass[10pt,a4paper]{iopart}
\usepackage{iopams}  
\usepackage{graphicx,psfrag,bbm,latexsym,color,dcolumn,bm,dsfont,bbm,
color,mathrsfs,bbold,latexsym,amsfonts,amssymb}

\newcommand{\media}[1]{\left\langle #1 \right\rangle_N}

\newcommand{\mediac}[1]{\left\langle #1 \right\rangle_{N}^{(c)}}
\newcommand{\mediap}[1]{\left\langle #1 \right\rangle_p}

\begin{document}
\title[Exact ground state for a class of matrix Hamiltonian models]
{Exact ground state for a class of matrix Hamiltonian models:
quantum phase transition and universality in the thermodynamic limit}

\author{Massimo Ostilli$^{1,2}$ and Carlo Presilla$^{1,2,3}$}

\address{$^1$\ Dipartimento di Fisica, Universit\`a di Roma ``La Sapienza'', 
Piazzale A. Moro 2, Roma 00185, Italy}
\address{$^2$\ Center for Statistical Mechanics and Complexity, 
INFM-CNR, Unit\`a di Roma 1, Roma 00185, Italy}
\address{$^3$\ Istituto Nazionale di Fisica Nucleare, Sezione di Roma 1, 
Roma 00185, Italy}

\date{\today}

\begin{abstract}
By using a recently proposed probabilistic approach,
we determine the exact ground state of a class of matrix Hamiltonian models
characterized by the fact that in the thermodynamic limit 
the multiplicities of the potential values assumed by the system
during its evolution are distributed according to 
a multinomial probability density.
The class includes 
\textit{i}) the uniformly fully connected models, namely a collection of 
states all connected with equal hopping coefficients and in the presence of 
a potential operator with arbitrary levels and degeneracies, and \textit{ii})
the random potential systems, in which the hopping operator is generic
and arbitrary potential levels are assigned randomly to the states 
with arbitrary probabilities.
For this class of models we find a universal thermodynamic limit 
characterized only by the levels of the potential, 
rescaled by the ground-state energy of the system for zero potential, 
and by the corresponding degeneracies (probabilities).
If the degeneracy (probability) of the lowest potential level tends to zero, 
the ground state of the system undergoes a quantum phase transition 
between a normal phase and a frozen phase with zero hopping energy.
In the frozen phase the ground state condensates into the subspace spanned 
by the states of the system associated with the lowest potential level.
\end{abstract}

\pacs{02.50.-r, 05.40.-a, 71.10.Fd}

\maketitle

\section{Introduction}
A multitude of evolution problems can be cast in the form of linear flows,
$\partial_t \psi = \hat{H} \psi$, 
where $\hat{H}$ is a matrix operator not necessarily Hamiltonian. 
For an initial value $\psi(0)$, the solution $\psi(t)$ of these systems 
of linear differential equations requires, as it is well known, 
the evaluation of the exponential of the operator $\hat{H}$.
The solution, with a real or an imaginary time $t$, also admits 
an exact probabilistic representation in terms of a proper collection 
of independent Poisson processes~\cite{DAJLS,DJS,BPDAJL}.

Recently, we have exploited this probabilistic representation to derive 
exact finite-time solutions of a Hamiltonian flow by a Monte Carlo 
algorithm~\cite{OP3},
as well as analytical results in the long-time limit~\cite{OP1,OP2}.
In the latter approach,
the Hamiltonian operator is decomposed, in a chosen representation, 
into diagonal and off diagonal parts, the potential and the hopping terms,
respectively. 
On applying a central limit theorem to the rescaled multiplicities  
of the values assumed by the potential and hopping terms in the configurations
visited by the system during its evolution,
we have obtained a simple scalar equation whose solution provides 
an approximated semi-analytical expression for the lowest 
eigenvalue of $\hat{H}$, in the following often called ground-state energy.

The scalar equation derived in~\cite{OP1} contains only the first 
two statistical moments of the potential and hopping multiplicities 
and suggests that it is a second order truncation of an exact 
cumulant expansion.
We have obtained this exact cumulant expansion by a large deviation 
analysis of the relevant probability density~\cite{OP4}.
In principle, were all the cumulants known, 
we would be in possession of a scalar equation whose straightforward 
solution is the exact lowest eigenvalue of $\hat{H}$.
In general, however, only a finite number of cumulants is available so that 
we have an approximated truncated equation whose level of accuracy 
depends on the system considered.
In some cases, the convergence as a function of the number of cumulants 
is rather fast and the use of the first few cumulants provides results 
indistinguishable from those obtained by exact 
numerical simulations, see \cite{OP4} for details.

In this paper we revisit the probabilistic approach~\cite{OP1,OP2,OP4} from
a different point of view. 
We suppose that the asymptotic probability density of the potential 
and hopping multiplicities is known and derive an exact equation 
for the ground state in a closed form, 
\textit{i.e.} not as a series expansion in cumulants. 
This formal result is then made concrete by observing that there exists
a class of systems whose associated probability density is a multinomial.
In fact, in the case of uniformly fully connected models and 
for random potential systems we obtain a very simple equation 
which relates the lowest eigenvalue of $\hat{H}$, $E_0$,  
to that of the Hamiltonian operator with zero potential, $E_0^{(0)}$.
This equation becomes exact in the limit $M \to \infty$, where $M$ is the 
size of the matrix representing $\hat{H}$, provided that in the same limit 
$E_0^{(0)}$ diverges, namely a thermodynamic limit.

By reason of the analytical expression found for $E_0(M)$ 
as a function of $E_0^{(0)}(M)$, supposed known,
the thermodynamic limit of the ground state of the above systems
can be exactly characterized.
We realize that, in this limit, a singularity may show up in the solution
for $E_0$ and a quantum phase transition takes place.
Such a behavior is obtained when the degeneracy, or probability,
the name depends on the kind of system we consider, 
of the lowest potential level vanishes.
We also provide the equation of the critical surface separating 
the two phases and show that the hopping energy represents an effective
order parameter of the transition.
For both the uniformly fully connected models and the random potential systems,
the thermodynamic features of the ground state, 
including the existence of a quantum phase transition, 
are universal, in the sense that these depend, 
up to a rescaling by $|E_0^{(0)}|$,
only on the levels and on the corresponding degeneracies (probabilities) 
of the potential.
In particular, for the random potential systems we obtain a universal
behavior independently of the nature of the hopping operator.
This conclusion compares rather well with the results of exact numerical 
solutions for finite-size systems of quantum particles in one- and 
two-dimensional lattices.
In fact, independently of the range chosen for the hopping operator, 
long or first neighbor, 
and independently of the nature of the particles, 
hard-core bosons or fermions, in all cases 
we have a clear tendency toward the universal thermodynamic behavior 
predicted by our formula.

The paper is organized as follows. 
In Section \ref{ufcm&rps} we introduce the uniformly fully connected models 
and the systems with random potential we deal with in the rest of the paper.
In Section \ref{outlinef} we summarize our main result, namely
a scalar equation whose solution provides the exact ground-state energy
of the Hamiltonian operator for the above models in the limit $M\to \infty$.
The proof of this result is given in Section \ref{proof}.
In Subsection \ref{apa} we revisit the probabilistic approach in
the case in which the asymptotic probability density of the potential 
and hopping multiplicities is known.
The lowest eigenvalue of $\hat{H}$ is obtained,
together with the asymptotic frequencies associated with the potential and 
hopping multiplicities, as the solution of a general system of equations.
This result is specialized to the case of
uniformly fully connected models and to random potential systems
in Subsections \ref{ufcm} and \ref{rps}, respectively.
In Section \ref{thermolimit} we discuss the thermodynamic limit 
and show its universal behavior and the appearing of a quantum phase 
transition when the degeneracy (probability) of the lowest potential level 
vanishes.
Comparisons with exact numerical results obtained for finite-size
lattice systems are provided in Section \ref{many-body.discrete} in 
the case of random potentials with discrete spectrum and in
Section \ref{many-body.continuous} for a potential with continuous spectrum.
The general features and results of our approach 
are summarized in Section \ref{conclusions}.

\section{Uniformly fully connected models and random potential systems}
\label{ufcm&rps}
Let us consider a finite set of $M$ states labeled by a vector index $\bm{n}$.
We will indicate this set as $\mathbb{F}$, 
\textit{i.e.} $\mathbb{F}=\{\bm{n}\}$ and $|\mathbb{F}|=M$.
The nature of the vector $\bm{n}$ depends on the context;
in the case of an Hamiltonian particle model, for example, $\bm{n}$ 
represents an element of a Fock space.
Let us also consider a state function
$V:\mathbb{F} \to \mathbb{R}$ with $V(\bm{n})=V_{\bm{n}}$.
We look for the lowest eigenvalue, $E_0$, of the $M\times M$ matrix $\bm{H}$
whose off-diagonal matrix elements are all equal to $-\eta$, 
with $\eta >0$ arbitrary, 
whereas the diagonal terms are given by the values $V_{\bm{n}}$, 
with $\bm{n} \in \mathbb{F}$. 
In equations,
\begin{eqnarray}
\label{Hf} 
\bm{H}= \bm{K} + \bm{V},
\end{eqnarray}
where the elements of the hopping and potential matrices, 
$\bm{K}$ and $\bm{V}$, are given by
\begin{eqnarray}
\label{ef}
K_{\bm{n},\bm{n}'} = -\eta(1-\delta_{\bm{n},\bm{n}'}) ,
\end{eqnarray}
\begin{eqnarray}
\label{Vf} 
V_{\bm{n},\bm{n}'} = V_{\bm{n}}\delta_{\bm{n},\bm{n}'} .
\end{eqnarray}
In the following, we will indicate with $\mathscr{V}$ 
the set of all the possible different values of 
$V_{\bm{n}}$ with $\bm{n} \in \mathbb{F}$,
the levels of the potential for brevity, 
and with $p_V$ the degeneracy of level $V$,
\begin{eqnarray}
\label{Vdf} 
p_V = \frac{1}{M} \sum_{\bm{n}} \delta_{V,V_{\bm{n}}} .
\end{eqnarray}
Note that $|\mathscr{V}|\leq M$ and $\sum_{V\in\mathscr{V}} p_V = 1$.

The matrix $\bm{H}$ is the $\mathbb{F}$ representation 
of a Hamiltonian operator $\hat{H}$ describing a fully connected model,
or complete graph.
This can be understood as follows.
In general, a linear operator $\hat{H}$ defines the time evolution of 
the state function $\psi(\bm{n};t)$ of the system according to the equation 
$\partial_t \psi(\bm{n};t)=\hat{H}\psi(\bm{n};t)$ 
starting from some given initial condition $\psi(\bm{n};0)$.  
We can always split $\hat{H}$ in two terms, 
$\hat{K}$ and $\hat{V}$, 
such that in the $\mathbb{F}$ representation the corresponding matrices have, 
respectively, only off-diagonal and diagonal non vanishing elements. 
As it is clear from the probabilistic representation of the evolution 
equation, see \cite{BPDAJL}, the off-diagonal terms of $\bm{H}$ 
are the rates for the transitions $\bm{n}\rightarrow \bm{n}'$ 
between two different states,
whereas the diagonal terms represent weights associated with 
the permanence in the states of the system.
Therefore, Eqs. (\ref{Hf}-\ref{Vf}) describe a system in which 
all the possible $M(M-1)$ transitions $\bm{n} \to \bm{n}'$, 
with $\bm{n} \neq \bm{n}'$, take place with the same rate,
namely a uniformly fully connected model.

The other class of models which we will consider in this paper
are defined in the following way.
Given an arbitrary hopping matrix with elements $K_{\bm{n},\bm{n}'}$,
the values of the potential matrix $V_{\bm{n}}$ 
are assigned to the states $\bm{n}$ randomly with
the probability
\begin{eqnarray}
\label{PV} 
P(V_{\bm{n}}=V) = 
\cases{
p_V,  &  $V\in\mathscr{V}$,\\ 
0,    &  $V\notin\mathscr{V}$,\\
} 
\end{eqnarray}  
where, now,  $\mathscr{V}$ is the set of the chosen levels of the 
potential $V$ and $\{p_V\}$ the set of the corresponding given  
normalized probabilities,
\begin{eqnarray}
\label{pV} 
\sum_{V\in\mathscr{V}} p_V=1.
\end{eqnarray}
The choice of $\mathscr{V}$ and $\{p_V\}$ is arbitrary.

The two classes of models described above are, in a sense, 
complementary to each other. 
In fact, for the former we have the particular hopping 
operator (\ref{ef}), which allows uniformly random connections
among all the states, and an arbitrary potential operator, 
whereas for the latter we have an arbitrary hopping operator 
and random potential levels.

\section{Main result}
\label{outlinef}
Consider first the uniformly random connected models. 
For $V\equiv 0$, the lowest eigenvalue of (\ref{Hf}), 
which we call $E_{0}^{(0)}$, 
is trivially related to the lowest eigenvalue of the
unit matrix and is given by 
\begin{eqnarray}
\label{E0f} 
E_0^{(0)}=-\eta (M-1).
\end{eqnarray}
It coincides, up to the factor $-\eta$, with the number of states
connected to any state of the system.
Similarly, for $V$ constant, 
let say $V_{\bm{n}}=V_0$ for any $\bm{n}\in\mathbb{F}$,
we have $E_0 = -\eta(M-1)+V_0$. 
Note that, for the particular value $V_0=-\eta$, we get the obvious result
$E_0=-\eta M$.

In the case of an arbitrary potential operator $\hat{V}$, an expression for $E_0$ is 
not known. 
By using the probabilistic approach developed in~\cite{OP1,OP2,OP4},
in this paper we will show that, in the limit $M\to\infty$,
$E_0$ is the unique solution of the following equation
\begin{eqnarray}
\label{Ef} 
\sum_{V\in\mathscr{V}} \frac{p_V}{-E_0+V}=\frac{1}{\eta (M-1)}, 
\qquad E_0\leq V_{\mathrm{min}} ,
\end{eqnarray}
or, in terms of the noninteracting energy $E_0^{(0)}$,
\begin{eqnarray}
\label{E1f} 
\sum_{V\in\mathscr{V}}
 \frac{p_V}{E_0-V}=\frac{1}{E_0^{(0)}}, 
\qquad E_0\leq V_{\mathrm{min}}.
\end{eqnarray}
It is easy to check that Eq. (\ref{E1f}) with $E_0$ as an unknown 
has always solution
and that the condition $E_0\leq V_{\mathrm{min}}$,
where $V_{\mathrm{min}}$ is the smallest element of $\mathscr{V}$,  
ensures the uniqueness of this solution.

Similarly, we will show that for $M\to\infty$ 
the ground state of the random potential systems defined by Eqs. 
(\ref{PV}) and (\ref{pV}) is also given by Eq. (\ref{E1f}),
provided that in the same limit $E_0^{(0)}$ diverges.
For these systems, the noninteracting ground-state energy
has no more a trivial form like in (\ref{E0f}), 
as it is the lowest eigenvalue of an arbitrary hopping matrix $\bm{K}$.
In this case, $E_0^{(0)}$ must be determined as a separated problem, 
\textit{e.g.} by using Monte Carlo simulations in the absence 
of sign problem, or by Fourier transformation if $\bm{K}$ represents 
the hopping operator of a system of independent particles.

It is worth to stress that even if, in general, $E_0$ is given only 
implicitly by Eq. (\ref{E1f}), this is a simple one-dimensional equation
which can be straightforwardly solved by an iterative method.

As we will show in details in Section 5, in the thermodynamic limit
the solution of Eq. (\ref{E1f}) may develop a singularity in its derivatives
with respect to the potential levels $V$, signaling a two-state quantum phase 
transition with a frozen phase having zero hopping energy.
These results hold for potentials with both discrete and continuous spectrum.

\section{Proof of Equation (\ref{E1f})}
\label{proof}
The derivation of Eq. (\ref{E1f}) follows from 
an analytical probabilistic approach we have recently developed 
to analyze the ground state of an arbitrary Hamiltonian 
operator $\hat{H}$ represented by a finite matrix.
In subsection \ref{apa} we review the basic definitions and results
of the above approach referring the reader to \cite{OP4} for details.
Although in \cite{OP4} we have explicitly considered lattice systems,
\textit{i.e.} the states $\bm{n}$ are lattice configurations with
$\bm{n}$ indicating the occupation numbers of the sites, 
nothing changes if one looks at the states $\bm{n}$ as arbitrary abstract
states. 
Moreover, differently from \cite{OP4}, we will assume that the
probability density which is the core of our approach is known.
As a result, we will find a formal equation for $E_0$ in a closed form.

In general, given an arbitrary Hamiltonian matrix $\bm{H}$, 
each row $\bm{n}$ of the corresponding hopping matrix $\bm{K}$ 
has a different number of non zero elements, let say $A(\bm{n})$. 
We call $A(\bm{n})$ the number of active links of the state $\bm{n}$.
When the number of active links of each state are all equal to the maximum 
value allowed, namely $A(\bm{n}) =  M-1$, 
we recover a fully connected model. 
First, we will consider the analytical probabilistic approach
in the general case. 
Applications to the uniformly fully connected models and 
to the random potential systems will be discussed in 
subsections \ref{ufcm} and \ref{rps}, respectively.

\subsection{Analytical probabilistic approach}
\label{apa}
Given the Hamiltonian operator $\hat{H}$ and separated 
its corresponding $\mathbb{F}$-representation matrix $\bm{H}$ 
into the hopping and potential matrices, $\bm{K}$ and $\bm{V}$, 
respectively, we define a virtual dynamics as follows.
Let us parametrize the matrix $\bm{K}$ as
\begin{eqnarray}
\label{Kf} 
K_{\bm{n},\bm{n}'}=\lambda_{\bm{n},\bm{n}'} ~\eta_{\bm{n},\bm{n}'}, 
\end{eqnarray}
such that $|\lambda_{\bm{n},\bm{n}'}|$ can be either 0 or 1  
and $\eta_{\bm{n},\bm{n}'}>0$.
In graph theory the matrix with elements $|\lambda_{\bm{n},\bm{n}'}|$ is
known as the adjacency matrix: in fact, it establishes whether
two given states $\bm{n},\bm{n}'$ are first neighbors or not.
We consider the Markov chain defined by
the transition matrix $\bm{P}$ with elements
\begin{eqnarray}
\label{Pf} 
P_{\bm{n},\bm{n}'}=\frac{|\lambda_{\bm{n},\bm{n}'}|}{A(\bm{n})},
\end{eqnarray}
where 
\begin{eqnarray}
\label{Af} 
A(\bm{n}) = \sum_{\bm{n}'} |\lambda_{\bm{n},\bm{n}'}| .
\end{eqnarray}
Starting from a given initial configuration $\bm{n}_0$, 
we draw a new configuration $\bm{n}_1$ with probability 
$P_{\bm{n}_0,\bm{n}_1}$.
By iterating this procedure for $N$ steps 
we construct a path, or trajectory, in the space $\mathbb{F}$
$\bm{n}_0,\bm{n}_1,\ldots,\bm{n}_N$. 
For simplicity, in this paper we consider $\hat{H}$ to be a Hamiltonian 
operator, \textit{i.e.} $\bm{H}$ is a complex Hermitian 
or real symmetric matrix. 
The approach can be generalized to non Hamiltonian operators.

We will show that the information 
about the ground state of $\bm{H}$ is contained in the ensemble of the 
infinitely long paths. 
Along each finite path with $N$ steps we have the sequences of data
$A_0,A_1,\ldots,A_N$, $V_0,V_1,\ldots,V_N$,
$\lambda_1,\ldots,\lambda_N$, 
and $\eta_1,\ldots,\eta_N$, where
\begin{eqnarray}
\label{Lf} 
A_k=A(\bm{n}_k), \qquad k=0,\dots,N,
\\
V_k=V(\bm{n}_k), \qquad k=0,\dots,N,
\\
\lambda_k=\lambda_{\bm{n}_{k-1},\bm{n}_k}, \qquad k=1,\dots,N,
\\
\eta_k=\eta_{\bm{n}_{k-1},\bm{n}_k}, \qquad k=1,\dots,N.
\end{eqnarray} 
For later use we also define the sequence of values
$T_0,T_1,\ldots,T_{N-1}$, where
\begin{eqnarray}
\label{Tf} 
T_k=A_k\eta_{k+1}/\epsilon, \qquad k=0,\dots,N-1,
\end{eqnarray} 
$\epsilon$ being an arbitrary reference constant, which has the 
same dimensions of the $\eta$'s, typically an energy.
Note that by construction $|\lambda_k|=1$.
Let us indicate with
$\mathscr{V}$, $\mathscr{T}$ and $\mathscr{L}$ the
sets of all the possible different values that can be taken by
the functions $V$, $T$ and $\lambda$, respectively.
Let $m_\mathscr{V}$, $m_\mathscr{T}$ and $m_\mathscr{L}$ 
be the cardinalities of these sets.
In \cite{OP4} we have shown that the ground state energy of $\bm{H}$ 
can be expressed in terms, not of the detailed sequences of the functions 
$V$, $T$ and $\lambda$, but just of the ensemble of 
their multiplicities, \textit{i.e.}
the number of times, $N_V$, $N_T$ and $N_{\lambda}$, 
a given value for $V$, $T$ and $\lambda$ has, respectively, 
occurred along each path, \textit{i.e.} 
\begin{eqnarray}
\label{Mf} 
N_V = \sum_{k=0}^{N}\delta_{V,V_k}, \\ 
N_T = \sum_{k=0}^{N-1}\delta_{T,T_k}, \\
N_{\lambda} = \sum_{k=0}^{N-1}\delta_{\lambda,\lambda_{k}}.
\end{eqnarray} 
Note that these multiplicities are normalized to the number of steps $N$
\begin{eqnarray}
\label{M1f} 
\sum_{V \in \mathscr{V}} N_V=N+1,\qquad 
\sum_{T \in \mathscr{T}} N_T =
\sum_{\lambda \in \mathscr{L}} N_{\lambda}=N.
\end{eqnarray} 
More precisely, in \cite{OP4} we have proven that the matrix elements of the 
evolution operator at time $t$ have the following probabilistic representation
\begin{eqnarray}
\label{Emf}
\sum_{\bm{n}} \langle \bm{n}|e^{-\hat{H}t} | \bm{n}_0\rangle =  
\sum_{N=0}^{\infty}
\media{
\mathcal{W}_N(\{N_V\};t)
\prod_{T \in \mathscr{T}} T^{N_T}\prod_{\lambda \in \mathscr{L}} 
{\lambda}^{N_{\lambda}}},
\end{eqnarray}
where $\media{\cdot}$ means an average over the paths
of length $N$ generated by the Markov chain (\ref{Pf}) starting from
the initial configuration $\bm{n}_0$.
These averages over paths of fixed length are named canonical.
The path functional weights $\mathcal{W}_N$ that appear in Eq. (\ref{Emf})
are defined by the system
\numparts
\begin{eqnarray}
\mathcal{W}_{N}(\{N_V\};t) =
\frac{e^{x_{0}t-\sum_{V\in\mathscr{V}} N_{V} \log[(x_{0}+V)/\epsilon]}}
{\sqrt{ 2\pi \sum_{V\in\mathscr{V}}\frac{\epsilon^2 N_{V}}{(x_{0}+V)^{2}} } },
\label{WSADDLE1f}
\\ 
\sum_{V \in \mathscr{V}}\frac{N_{V}}{x_{0}+V}=t, 
\qquad x_{0}>-V_{\mathrm{min}}.
\label{WSADDLE2f}
\end{eqnarray}
\endnumparts
For brevity, in the following we will drop the dependence 
of the weights on the potential multiplicities.

Once we have access to the canonical averages
and the series in the r.h.s. of Eq. (\ref{Emf}) is summed,
we obtain the lowest eigenvalue of $\bm{H}$ as
\begin{eqnarray}
\label{E0relationf} 
E_0 
= \lim_{t \to \infty} -\partial_t \log 
\sum_{\bm{n}} \langle \bm{n}|e^{-\hat{H}t} | \bm{n}_0\rangle,
\end{eqnarray}
with $\bm{n}_0$ arbitrary provided that it has a non zero projection 
onto the ground state $|E_0\rangle$. 

To evaluate the canonical averages it is useful to introduce the frequencies 
$\nu^{}_V=N_V/N$, $V \in \mathscr{V}$, 
$\nu^{}_T=N_T/N$, $T \in \mathscr{T}$, and
$\nu^{}_{\lambda}=N_{\lambda}/N$, $\lambda \in \mathscr{L}$,
which for $N$ large become continuously distributed in the range
$[0,1]$ with the constraints 
\begin{eqnarray}
\label{CONSTRAINTSf}
\sum_{V \in \mathscr{V}} \nu^{}_V = \sum_{T\in \mathscr{T}} \nu^{}_T
=\sum_{\lambda\in \mathscr{L}} \nu^{}_{\lambda}=1. 
\end{eqnarray}
Note that, for $N$ large, we do not distinguish the different normalizations,
$N+1$ for $N_V$ and $N$ for $N_T$ and $N_{\lambda}$, respectively.
When possible, for the multiplicities and the frequencies
we will use a compact notation in terms of the vectors
$\bm{\mu}$ and $\bm{\nu}$, which have
$m=m_{\mathscr{V}}+m_{\mathscr{T}}+m_{\mathscr{L}}$ components 
indicated by a Greek index 
$\alpha \in \mathscr{H} = \mathscr{V} \cup \mathscr{T}\cup \mathscr{L}$
and are defined as
$\bm{\nu}^\mathrm{T} = 
(\ldots \nu^{}_V \ldots; \ldots \nu^{}_T \ldots ; 
\ldots \nu^{}_{\lambda} \ldots)$.
We have
\begin{eqnarray}
\label{MUf}
\bm{\mu}=N\bm{\nu}.
\end{eqnarray}
For later use, we also define 
$\bm{u}^\mathrm{T}= 
(\ldots -\log[(x_{0}+V)/\epsilon] \ldots;
\ldots \log T \ldots; \ldots \log \lambda \ldots)$,
$\bm{v}^\mathrm{T} = 
(\ldots (x_{0}+V)^{-1} \ldots; \ldots 0 \ldots; \ldots 0 \ldots)$
and 
$\bm{w}^\mathrm{T} = 
(\ldots (x_{0}+V)^{-2} \ldots; \ldots 0 \ldots; \ldots 0 \ldots)$.
Note that the vectors $\bm{u}$, $\bm{v}$ and $\bm{w}$ 
depend on $\bm{\nu}$ through $x_{0}=x_{0}(\bm{\nu})$ and
$\bm{v}=-\partial_{x_{0}}\bm{u}$, $\bm{w}=-\partial_{x_{0}}\bm{v}$.
Finally, we will take advantage of a scalar product notation.
For instance, we rewrite Eq.~(\ref{WSADDLE2f}), which determines $x_0$, 
as $(\bm{\nu},\bm{v})=t/N$. 

By using the above compact notation, 
we express the $N$-th term of the series in the r.h.s. of Eq.~(\ref{Emf}) 
in the following explicit form
\begin{eqnarray}
\label{AVERAGE1f}
\media{ \mathcal{W}_{N}(t) \! \prod_{T\in \mathscr{T}} T^{N_T} 
\prod_{\lambda \in \mathscr{L} }\lambda^{N_{\lambda}}}  
= \sum_{\bm{\mu}}
\mathcal{P}_N(\bm{\mu}) 
\frac{e^{x_{0}t + \left( \bm{\mu},\bm{u} \right) }}
 { \sqrt {2\pi \epsilon^2 (\bm{\mu},\bm{w}) }},
\end{eqnarray}
where the probability $\mathcal{P}_N(\bm{\mu})$ is
given by the fraction of trajectories 
branching from the initial configuration $\bm{n}_0$ and having,
after $N$ steps, multiplicities $\bm{\mu}$.

In the limit of long times $t$ the paths most contributing to the 
evolution of the system are those with $N \sim t$ large, 
see \cite{OP4} for details. 
In this limit, therefore, we can change from discrete paths with 
multiplicity $\bm{\mu}$ to continuous paths with frequency $\bm{\nu}$  
\begin{eqnarray}\fl
\label{AVERAGE2f}
\media{ \mathcal{W}_{N}(t) \! \prod_{T\in \mathscr{T}} T^{N_T} 
\prod_{\lambda \in \mathscr{L}}\lambda^{N_{\lambda}}}&=& 
\int d(N\bm{\nu}) \mathcal{P}_N(N\bm{\nu}) 
\frac{ e^{x_{0}t + N\left( \bm{\nu},\bm{u} \right) } }
{ \sqrt {2\pi N \epsilon^2 (\bm{\nu},\bm{w}) }} 
\nonumber\\&=&
\int d(N\bm{\nu})  
\frac{ e^{N\left[ x_{0}t/N + \left( \bm{\nu},\bm{u} \right) 
+N^{-1} \log\mathcal{P}_N(N\bm{\nu}) \right] } }
{ \sqrt {2\pi N \epsilon^2 (\bm{\nu},\bm{w}) }} .
\end{eqnarray}
Here, we assume that the constraints (\ref{CONSTRAINTSf}) are automatically 
taken into account by the probability $\mathcal{P}_N(N\bm{\nu})$. 
Later, we will find more convenient to relax this feature of 
$\mathcal{P}_N(N\bm{\nu})$ 
and explicitly introduce proper Lagrange multipliers. 
For $t$ and $N$ large, with $N \sim t$, we can evaluate the 
integrals in Eq. (\ref{AVERAGE2f}) by steepest descent 
\begin{eqnarray}\fl
\label{AVERAGE2fsd}
\media{ \mathcal{W}_{N}(t) \! \prod_{T\in \mathscr{T}} T^{N_T} 
\prod_{\lambda \in \mathscr{L}}\lambda^{N_{\lambda}}} \simeq 
e^{x_{0}^{\mathrm{sp}}t + 
N\left[ \left( \bm{\nu}^{\mathrm{sp}},\bm{u}^{\mathrm{sp}} \right) + 
N^{-1} \log \mathcal{P}_N(N\bm{\nu}^{\mathrm{sp}})
\right]},
\end{eqnarray}
where $\bm{\nu}^\mathrm{sp}$ is the saddle point of the non-smooth, 
exponentially-varying part of the integrand in Eq. (\ref{AVERAGE2f})
and we added the superscript $^{\mathrm{sp}}$ to any function 
of $\bm{\nu}$ to indicate the value of this function for
$\bm{\nu}=\bm{\nu}^\mathrm{sp}$.
The symbol $\simeq$ means asymptotic logarithm equality.
The saddle-point frequency $\bm{\nu}^\mathrm{sp}$
is the solution of the system of equations
\begin{equation}
\label{spnualpha}
u_{\alpha}(x_0(\bm{\nu})) + 
N^{-1} \partial_{\nu_{\alpha}}\log\mathcal{P}_N(N \bm{\nu})
= 0,
\qquad \alpha\in\mathscr{H},
\end{equation}
which are obtained by derivating the exponent in Eq. (\ref{AVERAGE2f}) 
with respect to $\nu_{\alpha}$ and 
using the property $\left( \bm{\nu},\bm{v} \right)=t/N$.

The series of the canonical averages in Eq. (\ref{Emf}) is easily summed by 
replacing the sum with an integral over $N$, 
which is asymptotically exact for $t\to\infty$, see \cite{OP4}, 
and computing the integral by steepest descent. 
The result coincides, in the sense of asymptotic logarithm equality,
with the r.h.s. of Eq. (\ref{AVERAGE2fsd}) evaluated at the 
saddle point $N^{\mathrm{sp}}(t)$ solution of the equation
\begin{eqnarray}
\label{speforN}
\left( \bm{\nu}^{\mathrm{sp}},\bm{u}^{\mathrm{sp}} \right) + 
N^{-1} \log \mathcal{P}_N(N\bm{\nu}^{\mathrm{sp}}) 
+ N \partial_N 
\left[  N^{-1} \log \mathcal{P}_N(N\bm{\nu}^{\mathrm{sp}}) \right]
\nonumber\\ 
+ \frac{d x_{0}^{\mathrm{sp}}}{dN}
\left[ t - N \left( \bm{\nu}^{\mathrm{sp}},\bm{v}^{\mathrm{sp}} \right)
\right]  
\nonumber\\
\left. + N \sum_{\alpha\in\mathcal{H}} \frac{d\nu_{\alpha}^{\mathrm{sp}} }{dN}
\left[ u_{\alpha}^{\mathrm{sp}} + N^{-1} \partial_{\nu_{\alpha}^{\mathrm{sp}}}
\log\mathcal{P}_N(N \bm{\nu}^{\mathrm{sp}}) 
\right]
\right|_{N=N^{\mathrm{sp}}(t)} = 0 .
\end{eqnarray}
The last two terms of the above equation take into account
the non explicit $N$-dependence of the exponent in Eq. (\ref{AVERAGE2fsd})
through $x_0^{\mathrm{sp}}$ and $\bm{\nu}^{\mathrm{sp}}$, respectively.
We get rid of the third term in Eq. (\ref{speforN})
by observing that for $N$ large we have 
\begin{eqnarray}
\label{OMEGA}
\lim_{N\to\infty} N^{-1}\log \mathcal{P}_N(N\bm{\nu}) = \omega(\bm{\nu}),
\end{eqnarray}
where $\omega(\bm{\nu})$ depends only on the frequencies $\bm{\nu}$, 
see \ref{Appendice}.
The fourth and fifth terms in Eq. (\ref{speforN}) also vanish due
to the property
$\left( \bm{\nu}^{\mathrm{sp}},\bm{v}^{\mathrm{sp}} \right) =t/N$
and the saddle-point equation for $\bm{\nu}^{\mathrm{sp}}$,
respectively. 
In conclusion, $N^{\mathrm{sp}}(t)$ is determined by
\begin{eqnarray}
\label{speforN2}
\left.
\left( \bm{\nu}^{\mathrm{sp}},\bm{u}^{\mathrm{sp}} \right) + 
N^{-1} \log \mathcal{P}_N(N\bm{\nu}^{\mathrm{sp}}) 
\right|_{N=N^{\mathrm{sp}}(t)} = 0.
\end{eqnarray}
As expected from a physical point of view, we have
$N^{\mathrm{sp}}(t)\sim t$ when $t\to \infty$, 
see \cite{OP4} for details.
Finally, we obtain
\begin{eqnarray}
\label{Emf1}
\sum_{\bm{n}} \langle \bm{n}|e^{-\hat{H}t} | \bm{n}_0\rangle 
\simeq 
\left. e^{ x_{0}^{\mathrm{sp}} t} \right|_{N=N^{\mathrm{sp}}(t)}.
\end{eqnarray}
Observing that for $t$ large we must have 
$\sum_{\bm{n}} \langle \bm{n}|e^{-\hat{H}t} | \bm{n}_0\rangle 
\simeq 
e^{ -E_0 t}$,
the ground-state energy $E_0$ is given by 
\begin{eqnarray}
\label{E0}
E_{0} = - \lim_{t \to \infty}~
\left. x_{0}^{\mathrm{sp}} \right|_{N=N^{\mathrm{sp}}(t)}. 
\end{eqnarray}
Taking into account the definition of $\bm{u}$, which
depends on $x_0$, in the limit $t \to \infty$
Eq. (\ref{speforN2}) can be read as a a time independent equation 
that determines the quantity $E_0$.
Note that in the same limit
$\left. \bm{\nu}^{\mathrm{sp}} \right|_{N=N^{\mathrm{sp}}(t)}$
assumes a constant value.

The key ingredient of the approach outlined above 
is the knowledge of the probability $\mathcal{P}_N(\bm{\mu})$.
Once this is known, the ground state can be derived
analytically by solving the equation
\begin{eqnarray}
\label{speforN3}
\lim_{t\to\infty}
\left[
\left( \bm{\nu}^{\mathrm{sp}},\bm{u}^{\mathrm{sp}} \right) + 
N^{-1} \log \mathcal{P}_N(N\bm{\nu}^{\mathrm{sp}}) 
\right]_{N=N^{\mathrm{sp}}(t)} = 0,
\end{eqnarray}
where
$\lim_{t\to\infty} 
\left. \bm{\nu}^{\mathrm{sp}} \right|_{N=N^{\mathrm{sp}}(t)}$
is a finite vector dependent on $E_0$ and 
\begin{eqnarray}\fl
\label{ulim}
\lim_{t\to\infty} 
\left( \left. \bm{u}^{\mathrm{sp}} \right|_{N=N^{\mathrm{sp}}(t)} 
\right)^\mathrm{T}
= 
(\ldots -\log[(-E_0+V)/\epsilon] \ldots;
\ldots \log T \ldots; \ldots \log \lambda \ldots).
\end{eqnarray}
  
Equation (\ref{speforN3}) allows the determination of the ground-state 
energy $E_0$ in terms of asymptotic saddle-point frequencies, 
for brevity $\bm{\nu}$ hereafter, 
which are determined by the system (\ref{spnualpha})
in the limit $N\rightarrow \infty$.
Therefore, by using again Eq. (\ref{OMEGA}), we find that
$E_0$ and $\bm{\nu}$ are the solution of
the following system of coupled equations
\begin{equation}
\label{SYSTEM}
\left\{\eqalign{
u_\alpha(E_0)+\partial_{\nu_\alpha} \omega(\bm{\nu})+c_\alpha =0,
\qquad \alpha\in\mathscr{H},
\\ 
\left(\bm{\nu},\bm{u}(E_0)\right) + \omega(\bm{\nu})=0,
\qquad E_0 \leq V_\mathrm{min},
\\
\sum_{V\in\mathscr{V}} \nu_V =1, 
\\
\sum_{T\in\mathscr{T}} \nu_T =1,
\\
\sum_{\lambda\in\mathscr{L}} \nu_\lambda =1,
}\right.
\end{equation}
where
$\bm{u}^\mathrm{T} = 
(\ldots -\log[(-E_0+V)/\epsilon] \ldots;
\ldots \log T \ldots; \ldots \log \lambda \ldots)$
and 
\begin{eqnarray}
c_\alpha =
\cases{
c_\mathscr{V}, & $\alpha\in\mathscr{V}$,\\ 
c_\mathscr{T}, & $\alpha\in\mathscr{T}$,\\
c_\mathscr{L}, & $\alpha\in\mathscr{L}$,\\
} 
\end{eqnarray}
are the three Lagrange multipliers introduced to explicitly take into account
the constraints (\ref{CONSTRAINTSf}) so that the function 
$\omega(\bm{\nu})$, \textit{i.e.} the corresponding probability, 
is now meant to have support on the whole hypercube $[0,1]^m$. 
In other words, the function $\omega(\bm{\nu})$ in the system (\ref{SYSTEM}) 
is the analytic continuation to the set $[0,1]^m$ of the function defined in 
Eq. (\ref{OMEGA}).

In \cite{OP4} we have used a cumulant expansion theorem to express,
in the general case in which $\mathcal{P}_N(\bm{\mu})$ is not known, 
$E_0$ as a function of the cumulants of the variables $V$, $T$ and $\lambda$ 
generated the Markov chain over the states $\bm{n}$.
It is clear that in this case only the first few cumulants can be considered
available, numerically or analytically, and the expression found for $E_0$ is 
necessarily approximated.
On the other hand, the exact value of $E_0$ can be obtained in the cases 
in which $\mathcal{P}_N(\bm{\mu})$ is known. 
As we shall show in the next two subsections, 
the uniformly fully connected models and the random potential systems
provide non trivial and important examples of the latter class.

\subsection{Uniformly fully connected models}
\label{ufcm}
Up to now we have formulated the problem in the most general case.
In this Section,
we specialize the discussion to the uniformly fully connected models
defined by Eqs. (\ref{Hf}-\ref{Vf}). 
For these models, 
both the sets $\mathscr{T}$ and $\mathscr{L}$ have a single
element, namely $T=(M-1)\eta/\epsilon$ and $\lambda=1$, 
whereas we may count, in general, 
$M$ distinct values $V$ in the set $\mathscr{V}$.
This implies that we have to consider only the potential multiplicities $N_V$,
$V \in \mathscr{V}$, 
\textit{i.e.} $\bm{\mu}^\mathrm{T}= (\ldots N_V \ldots ;N;N)$.
Furthermore, due to the fact that all the transitions
$\bm{n}\to\bm{n}'$, with $\bm{n} \neq \bm{n}'$ are allowed with the 
same probability,
we can assume that after $N$ steps the Markov chain (\ref{Pf}) 
is well represented by a multinomial distribution, 
in which, at each step, any state has the probability $1/M$ to be extracted.
This representation will become exact in the limit $M\to\infty$, 
\textit{i.e.} when the introduced extra transitions $\bm{n}\to\bm{n}$
become immaterial.
According to this multinomial distribution, 
the probability to have, after $N$ steps, multiplicities $\bm{\mu}$ 
satisfying the constraint $\sum_{V\in \mathscr{V}} N_V = N$, 
is given by
\begin{eqnarray}
\label{MULTIf0}
\mathcal{P}_N(\bm{\mu})= N! \prod_{V\in\mathscr{V}} 
\frac{{p_{V}}^{N_V}}{N_V!}.
\end{eqnarray}
The parameters $p_V$, with $V \in \mathscr{V}$, are 
the degeneracies of the potential values $V$ 
in the space of the states $\mathbb{F}$, see Eq. (\ref{Vdf}).
Due to the ergodic properties of the Markov chain (\ref{Pf}), 
they also coincide with the asymptotically expected frequencies 
of the multiplicities $N_V$ 
\begin{eqnarray}
p_V = \lim_{N\to \infty} \frac{1}{N} \media{N_V} .
\end{eqnarray}
The latter expression is relevant to calculate by Monte Carlo simulations 
the parameters $p_V$ in those cases in which a direct evaluation 
in the space $\mathbb{F}$ is not feasible.

By using the Stirling approximation for $N$ large,
the probability for the potential multiplicities $N\bm{\nu}$, 
with each component of $\bm{\nu}$ continuously distributed 
in the range $[0,1]$, can be written as
\begin{eqnarray}
\label{MULTI}
\mathcal{P}_N(N\bm{\nu}) 
\simeq
\exp\left[ N \sum_{V\in\mathscr{V}} \nu_V 
\log\left(\frac{p_V}{\nu_V}\right) \right],
\end{eqnarray}
from which we get
\begin{eqnarray}
\label{OMEGAMULTI}
\omega(\bm{\nu}) =
\sum_{V\in\mathscr{V}} \nu_V \log\left(\frac{p_V}{\nu_V}\right).
\end{eqnarray}

Taking into account that for this model we have
$\nu_{T\equiv (M-1)\eta/\epsilon}= 1$ and $\nu_{\lambda \equiv 1}= 1$, 
the only random multiplicities being those of the potential values,
the system of Eqs. (\ref{SYSTEM}) becomes
\begin{equation}\fl
\label{MULTISYSTEM}
\left\{\eqalign{
u_V(E_0)+\log\left(\frac{p_V}{\nu_V}\right)-1+c_\mathscr{V}=0, 
\qquad V \in \mathscr{V},\\
\sum_{V\in\mathscr{V}} \nu_V u_V(E_0)+
\log\left[(M-1)\eta/\epsilon\right]+
\sum_{V\in\mathscr{V}} \nu_V \log\left(\frac{p_V}{\nu_V}\right)=0,
\qquad E_0 \leq V_\mathrm{min},\\
\sum_{V\in\mathscr{V}} \nu_V =1.\\
}\right.
\end{equation}
By using $u_V(E_0)=-\log[(-E_0+V)/\epsilon]$ and observing that 
the Lagrange multiplier $c_\mathscr{V}$ does not depend on the index $V$, 
from the first equation of the above system we find 
\begin{eqnarray}
\label{SP1}
\nu_{V}= \frac{1}{Z} ~\frac{p_V}{-E_0+V},
\end{eqnarray}
where the normalization constant, $Z=\epsilon^{-1} \exp(c_\mathscr{V}-1)$, 
is determined by using the third equation in (\ref{MULTISYSTEM}), namely
\begin{eqnarray}
\label{MULTIC}
Z= \sum_{V\in\mathscr{V}} \frac{p_V}{-E_0+V}.
\end{eqnarray}
By inserting the result (\ref{SP1}) into the second equation of the system 
(\ref{MULTISYSTEM}), we get
\begin{eqnarray}
\label{MULTIE}
\frac{1}{Z} \log(Z\epsilon)
\sum_{V\in\mathscr{V}} \frac{p_V}{-E_0+V}+
\log\left[(M-1){\eta}/\epsilon\right] =0,
\end{eqnarray} 
which, on using the value (\ref{MULTIC}) for $Z$, 
brings immediately to Eq. (\ref{Ef}).

\subsection{Random potential systems}
\label{rps}
The above derivation can be readily extended to those systems 
in which, both the hopping values $T$ and the phases $\lambda$ are general, 
whereas the potential levels $V\in\mathscr{V}$ are independent random 
variables assigned to the states $\bm{n}$ with arbitrary probabilities $p_V$, 
see Section \ref{ufcm&rps}.
It is clear that different realizations of the values $V_{\bm{n}}$
correspond to different Hamiltonian matrices, \textit{i.e.}
we deal with an ensemble of random matrices.
In this case, therefore, the study of the ground-state energy 
should be meant as 
the evaluation of the average lowest eigenvalue of the ensemble
and of its fluctuations around this value.
In the following, we will assume that 
the probability for the ground-state energy to assume a given 
value becomes infinitely peaked around its average $E_0$. 

For a finite system with a specific realization of the potential
values $V_{\bm{n}}$ the corresponding ground-state energy is affected by the 
correlations among the $V_{\bm{n}}$'s induced by the virtual hopping dynamics.
Consider, in particular, the possibility that a state $\bm{n}$ comes 
back to itself after a few steps.
However, such correlations must vanish if the number
of active links associated with each state, $A(\bm{n})$, diverges.
In this limit, any step of the Markov chain is equivalent to 
a random extraction of the value $V$ 
distributed according to the given $p_V$.
In fact, the situation in which the numbers $A(\bm{n})$ increase by 
increasing the number of states $M$ is rather common, later we will consider
the important example of many-body systems on a lattice.
For these systems,
in the limit $M\to\infty$ the potential becomes exactly 
uncorrelated from the states $\bm{n}$ and 
along the virtual dynamics there holds the following factorization
\begin{eqnarray}
\label{RANDOM}
\mathcal{P}_N(\{N_V\},\{N_T\},\{N_\lambda\})=
\mathcal{P}_N^{\mathscr{V}}(\{N_V\}) ~
\mathcal{P}_N^{(0)}(\{N_T\},\{N_\lambda\}),
\end{eqnarray} 
where the superscripts $^{\mathscr{V}}$ and $^{(0)}$
refer to the probability density for the potential multiplicities
$\{N_V\}$ and for the rest of the variables $\{N_T\}$ and 
$\{N_\lambda\}$, respectively.
In terms of the asymptotic function $\omega$ defined by Eq. (\ref{OMEGA}), 
this implies that
\begin{eqnarray}
\label{RANDOM1}
\omega(\bm{\nu})=\omega^{\mathscr{V}}(\{\nu_V\})+~
\omega^{(0)}(\{\nu_T\},\{\nu_\lambda\}).
\end{eqnarray} 

The consequences of the factorization (\ref{RANDOM}) are immediately obtained.
Note that in the probabilistic representation illustrated 
in Subsection \ref{apa}, unlike other probabilistic representations 
in which importance sampling is included, 
the frequencies $\nu_T$ and $\nu_{\lambda}$ are independent of the
potential values, so that their distribution can be evaluated for $V\equiv 0$.
In this case, we have $\nu_{V\equiv 0}=1$ and 
$u_{V\equiv 0}= -\log(-E_0^{(0)}/\epsilon)$ and, therefore,
\begin{eqnarray}
\label{RANDOM2}
(\bm{\nu},\bm{u}(E_0^{(0)}) )=
-\log(-E_0^{(0)}/\epsilon) + 
\sum_{T\in\mathscr{T}}\nu_T\log T + 
\sum_{\lambda\in\mathscr{L}}\nu_{\lambda}\log \lambda .
\end{eqnarray} 
From this relation and from the second equation of the system (\ref{SYSTEM}), 
we find that the energy of the noninteracting system, $E_0^{(0)}$,
satisfies the equation
\begin{eqnarray}
\label{RANDOM3}
\log(-E_0^{(0)}/\epsilon) = 
\sum_{T\in\mathscr{T}}\nu_T\log T +
\sum_{\lambda\in\mathscr{L}}\nu_{\lambda}\log \lambda
+\omega^{(0)}(\{\nu_T\},\{\nu_\lambda\}) .
\end{eqnarray} 
From Eq. (\ref{RANDOM1}) and assuming that $E_0^{(0)}$ is known, 
we can use Eq. (\ref{RANDOM3}) to get rid of the quantities in the r.h.s. 
in the general case $V \neq 0$.
In fact, by inserting Eqs. (\ref{RANDOM1}) and (\ref{RANDOM3}) 
in the general system (\ref{SYSTEM}),
we are left with the following reduced system 
\begin{equation}\fl
\label{RANDOMSYSTEM}
\left\{\eqalign{
u_V(E_0)+\partial_{\nu_V}\omega^{\mathscr{V}}(\{\nu_V\})
+c_{\mathscr{V}}=0, 
\qquad V\in\mathscr{V},
\\ 
\sum_{V\in\mathscr{V}} \nu_V u_V(E_0)+
\log( -E_0^{(0)}/\epsilon)+
\omega^{\mathscr{V}}(\{\nu_V\})=0,
\qquad E_0 \leq V_\mathrm{min},
\\
\sum_{V\in\mathscr{V}} \nu_V =1.
}\right.
\end{equation}
Finally, by using the fact that the potential levels $V\in\mathscr{V}$ 
are randomly assigned with probabilities $p_V$, 
we have that $\omega^{\mathscr{V}}(\{\nu_V\})$ is again given 
by the multinomial formula (\ref{OMEGAMULTI}). 
Therefore, 
apart from the exchange of $E_0^{(0)}$ with $-(M-1)\eta$,
the system (\ref{RANDOMSYSTEM}) is formally identical to the
system (\ref{MULTISYSTEM}) so that, in the limit $M\to \infty$,
the ground-state energy $E_0$ of the random potential systems satisfies 
Eq. (\ref{E1f}).

\section{Thermodynamic limit: quantum phase transition and universality}
\label{thermolimit}
For the uniformly fully connected models and the random potential systems, 
the solution $E_0(M)$ of Eq. (\ref{E1f}) approaches the exact 
ground-state energy when the number of states $M$ becomes infinitely large
provided that the number of active links associated with each state, 
$A(\bm{n})$, also diverges.
The condition on the number of active links, 
which is evidently fulfilled by any uniformly fully connected model, 
for random potential systems can be viewed as a condition on the 
noninteracting ground-state energy $E_0^{(0)}(M)$, which is proportional
to a proper average of $A(\bm{n})$ with $\bm{n} \in \mathbb{F}$. 
In this Section, we will study a non trivial thermodynamic 
$M\to\infty$ limit of Eq. (\ref{E1f})
always assuming that 
$\lim_{M\to \infty} |E_0^{(0)}(M)| =\infty$,
as happens for any system of particles with fixed density.

For simplicity, we suppose that the number of potential levels, 
\textit{i.e.} the cardinality $m_\mathscr{V}$ of the set $\mathscr{V}$,
does not change by increasing the number of states $M$. 
Consistently, we assume that the degeneracies (probabilities) $\{p_V(M)\}$
tend to constant values for $M\to\infty$.
Looking at Eq. (\ref{E1f}), it is immediately clear that if the potential
levels $\{V(M)\}$ diverge for $M\to\infty$ more slowly than $E_0^{(0)}(M)$,
we have $\lim_{M\to\infty} E_0(M)/E_0^{(0)}(M)=1$.
Another trivial limiting solution is found
if the potential levels $\{V(M)\}$ diverge faster than $E_0^{(0)}(M)$,
namely $\lim_{M\to\infty} E_0(M)/V_\mathrm{min}(M)=1$.
A thermodynamic non trivial limit is obtained only for $V(M) \sim E_0^{(0)}(M)$.
In this limit, we rewrite Eq. (\ref{E1f}) as 
\begin{eqnarray}
\label{thermo1}
\sum_{v\in\mathscr{V}}
 \frac{p_v}{v-e_0}=1,
\qquad e_0\leq v_{\mathrm{min}},
\end{eqnarray}
where we have defined 
$p_v=\lim_{M\to\infty} p_{V(M)}$,
$v=\lim_{M\to\infty} V(M)/|E_0^{(0)}(M)|$
and 
$e_0=\lim_{M\to\infty} E_0(M)/|E_0^{(0)}(M)|$
and we have assumed, as usual if $\hat{K}$ represents a hopping operator, 
that $E_0^{(0)}(M)<0$.
With an abuse of notation, we keep using the same symbol chosen for 
the set $\{V\}$ also for the set of the asymptotic rescaled 
potential levels $\{v\}$. 
The asymptotic rescaled ground-state energy $e_0$ cannot exceed the minimum 
asymptotic rescaled potential level 
$v_\mathrm{min} =\min_{v\in\mathscr{V}} \{v\}$.
In the following, we will find handy 
ordering the elements of $\mathscr{V}$ as 
$v_1<v_2<\ldots<v_{m_\mathscr{V}}$,
and indicate with $p_1,p_2,\ldots,p_{m_\mathscr{V}}$ the 
corresponding degeneracies (probabilities).

For what concerns the uniformly fully connected models,
although Eq. (\ref{thermo1}) is formally equivalent to Eq. (\ref{E1f}), 
it has a reacher structure.
In fact, unlike the finite $M$ case in which the $p_V$'s vary in the
rational field, see Eq. (\ref{Vdf}), in the limit $M\rightarrow \infty$
the $p_v$'s vary in the continuum so that we are 
allowed to consider the analytic continuation of the solution $e_0$ 
toward limits in which one or more of the $p_v$'s tend to zero.
Regarding the random potential systems a similar observation,
strictly speaking, does not apply.  
In this case, in fact, the $p_V$'s can vary in the continuum
even with a finite $M$, see Eq. (\ref{PV}). 
On the other hand, as explained in the previous Section, 
for the random potential systems the ground-state energy becomes
a well defined quantity with vanishing relative fluctuations only in
the limit $M\rightarrow \infty$. 

We shall show that, due to the constraint $e_0\leq v_\mathrm{min}$,
for $p_1\to 0$ the solution $e_0$ of  Eq. (\ref{thermo1}) becomes non analytic 
and a quantum phase transition takes place.

\begin{figure}[t]
\centering
\includegraphics[width=0.7\columnwidth,clip]{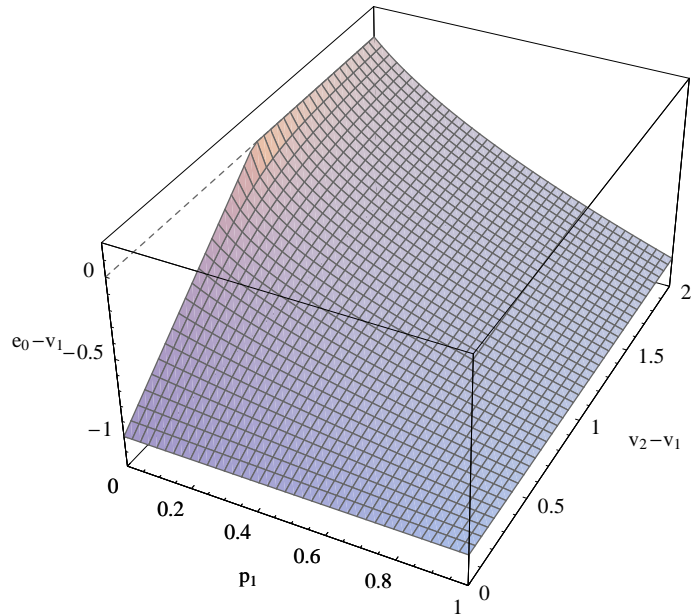}
\caption{ 
Thermodynamic limit of a system with two potential levels: 
asymptotic rescaled ground-state energy $e_0-v_1$ as a function of 
$p_1$ and $v_2-v_1$, see Eq. (\ref{thermo2}).}
\label{multinomial.fig1}
\end{figure}

\begin{figure}[t]
\centering
\includegraphics[width=0.7\columnwidth,clip]{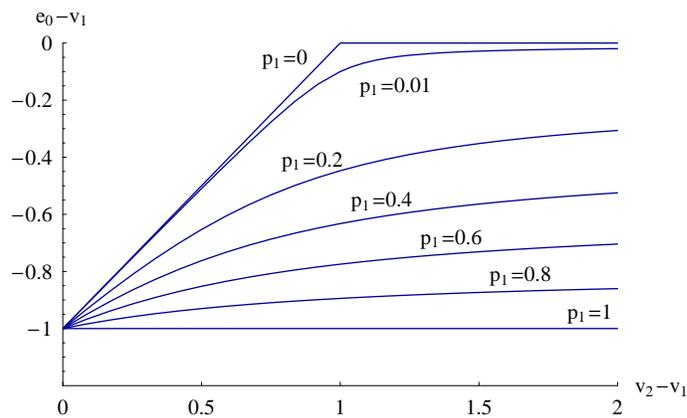}
\caption{
As in Fig. \ref{multinomial.fig1}: 
sections of $e_0-v_1$ as a function of $v_2-v_1$ for different 
values of $p_1$. For $p_1=0$ a quantum phase transition takes place at 
the critical point $v_2-v_1=1$.}
\label{multinomial.fig2}
\end{figure}

First we illustrate in detail an example with only two potential levels.
In this case, Eq. (\ref{thermo1}) is a quadratic equation for $e_0$
which can be solved explicitly.
Observing that $p_2=1-p_1$, we find
\begin{eqnarray}
\label{thermo2}
e_0= v_1 - \case{1}{2}
\left[
\sqrt{(v_2-v_1-1)^2 + 4 p_1 (v_2-v_1)}
-(v_2-v_1-1)
\right],
\end{eqnarray}
the other solution being incompatible with the condition $e_0\leq v_1$.
The behavior of $e_0-v_1$ as a function of $p_1$ and $v_2-v_1$ is 
shown in Fig. \ref{multinomial.fig1}.
For $p_1\to 0$ we have 
\begin{eqnarray}
\label{thermoe}
\lim_{p_1 \to 0}e_0-v_1 &=  
\case{1}{2} \left[ (v_2-v_1-1) - |v_2-v_1-1| \right]
\nonumber \\
&= \cases{
v_2-v_1-1, & $v_2-v_1 < 1$, \\
0,   & $v_2-v_1 > 1$, \\
}
\end{eqnarray}
\textit{i.e.} a singularity shows up at $v_2-v_1=1$, 
see Fig. \ref{multinomial.fig2}.
It is trivial to check that for $p_1=0$, $e_0$ has a discontinuity in
the first derivative and a divergence in the second derivative 
with respect to $v_2-v_1$ at $v_2-v_1=1$.
In conclusion, for $p_1=0$ we have a quantum phase transition
at the critical point $v_2-v_1=1$.
We stress that this behavior emerges due to the constraint $e_0\leq v_1$,
which keeps to hold only if the limits performed are taken in the correct 
order: $M\to\infty$ first and $p_1\to 0$ later.

\begin{figure}[t]
\centering
\includegraphics[width=0.99\columnwidth,clip]{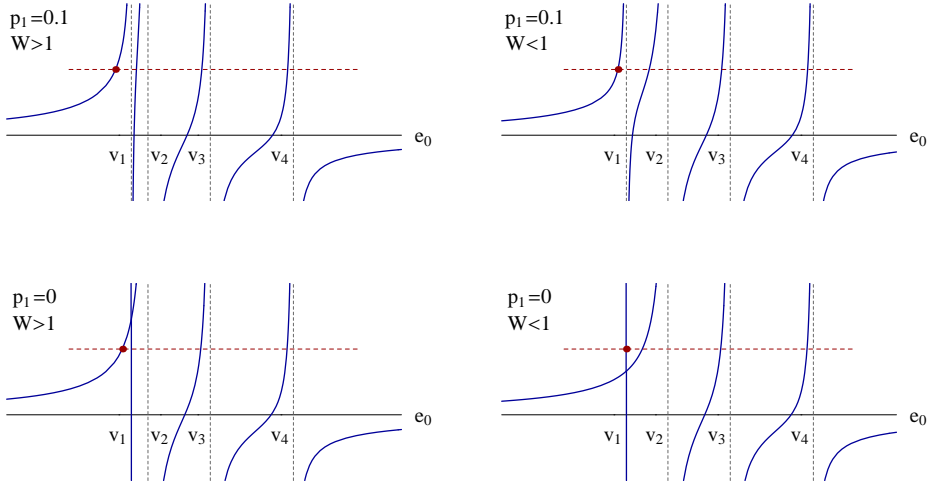}
\caption{Graphical solution of Eq. (\ref{thermo7}) in a system with 
four potential levels for $p_1=0.1$ (top) and $p_1=0$ (bottom)
in two cases $W>1$ (left) and $W<1$ (right).
The continuous lines are the l.h.s. of Eq. (\ref{thermo7}) 
plotted as a function of $e_0$ whereas  
the horizontal dashed lines represent the unit level.
The rescaled ground-state energy is given
by the unique intersection (dots) in the region $e_0\leq v_1$.
For $p_1=0$, a singularity shows up at $W=1$, 
namely $e_0$ stalls at its maximum value $e_0=v_1$ for $W\leq 1$.}
\label{multinomial.fig3}
\end{figure}

It is easy to generalize the above result to a case with $m_\mathscr{V}>2$.
Let us rewrite Eq. (\ref{thermo1}) as
\begin{eqnarray}
\label{thermo7}
-\frac{p_1}{e_0-v_1}+
\sum_{k=2}^{m_\mathscr{V}} \frac{p_k}{(v_k-v_1)-(e_0-v_1)}=1,
\qquad e_0-v_1 \leq 0.
\end{eqnarray}
The unknown $e_0-v_1$ is a function of the $2(m_\mathscr{V}-1)$ 
parameters $p_k,v_k-v_1$, $k=2,\ldots,m_\mathscr{V}$. 
Note that $p_1$ is fixed by the normalization condition.
We show the graphical construction of the solution of Eq. (\ref{thermo7})
in Fig. \ref{multinomial.fig3}. 
For $p_1>0$, \textit{i.e.} $\sum_{k=2}^{m_\mathscr{V}} p_k<1$, 
we always have $e_0<v_1$ and, furthermore, $e_0$ is an analytic function 
of its arguments.
When $p_1=0$, a singularity of $e_0$ shows up, which can be characterized 
in the following way.
Let us define the function
\begin{eqnarray}
\label{thermoW}
W(p_2,v_2-v_1,\ldots,p_{m_\mathscr{V}},v_{m_\mathscr{V}}-v_1) =
\sum_{k=2}^{m_\mathscr{V}} \frac{p_k}{v_k-v_1}.
\end{eqnarray}
From Eq. (\ref{thermo7}) we see that, if $p_1=0$ and
$e_0<v_1$, then 
\begin{equation}
  W > \sum_{k=2}^{m_\mathscr{V}} \frac{p_k}{(v_k-v_1)+(v_1-e_0)}=1.
\end{equation}
On the other hand, if $e_0 \to v_1$ for $p_1\to 0$, 
from the same Eq. (\ref{thermo7}) we have
\begin{eqnarray}
\label{thermoL}
1-W = \lim_{p_1\to 0, e_0 \to v_1^{-}}
\frac{p_1}{v_1-e_0} \geq 0,
\end{eqnarray}
\textit{i.e.} $W\leq 1$.
In conclusion, for $p_1\to 0$ the below relations hold between 
$W$ and $e_0$ 
\begin{eqnarray}
\label{thermo9}
W > 1 \Leftrightarrow  e_0<v_1, \\
\label{thermo9b}
W\leq 1 \Leftrightarrow  e_0=v_1.
\end{eqnarray}
These two equations establish the following scenario for $p_1\to 0$.
As we move inside the $2(m_\mathscr{V}-1)$ dimensional region 
determined by the condition $W\leq 1$, 
the rescaled energy $e_0$ stalls at its maximum value $e_0=v_1$. 
Outside this region, we have $e_0<v_1$. 
Therefore, we find that 
$\nabla e_0|_{\mathscr{S}^{+}}\neq 0$ and  
$\nabla e_0|_{\mathscr{S}^{-}}=0$,  
where $\mathscr{S}^{+}$ and $\mathscr{S}^{-}$
are generic points arbitrary close to the surface
$\mathscr{S}$, determined by the condition $W=1$, and 
such that $W>1$ or $W<1$, respectively.
More precisely, we find that on the critical surface $\mathscr{S}$
any directional derivative of $e_0$ has a discontinuity for
any direction not tangent to $\mathscr{S}$, which,
in turn, implies a divergence of any double derivative of $e_0$
for any direction not tangent to $\mathscr{S}$.

We observe that according to Eq. (\ref{thermoW}) 
the critical surface is determined by a sum over all the potential levels
of the ratios $p_k/(v_k-v_1)$. 
Therefore, even levels very far from the lowest one, $v_1$, can contribute
to the critical behavior if the corresponding degeneracies are large enough.
This is a cooperative phenomenon due to the intrinsic quantum nature
of the considered systems.

We wish to discuss now
in more detail the nature of the phase transition 
corresponding to the above singularity and also determine an order parameter. 
Let us write the potential operator $\hat{V}$ in terms of the 
projectors onto the configuration subspaces at fixed potential values
\begin{eqnarray}
\label{thermo10}
\hat{V}=\sum_{V\in\mathscr{V}} V \hat{\pi}_V,
\qquad
\hat{\pi}_V=\sum_{\bm{n}\in \mathbb{F}: V(\bm{n})=V} 
|\bm{n}\rangle\langle\bm{n}|.
\end{eqnarray}
On derivating the ground-state energy $E_0$ with respect to
a potential level $V$ and using the Hellman-Feynman theorem,
we find
\begin{eqnarray}
\label{thermo12}
\frac{\partial E_0}{\partial V}
=\frac{\partial}{\partial V}
\frac{\langle E_0 |\hat{H}| E_0 \rangle}
{\langle E_0 | E_0 \rangle}
=
\frac{\langle E_0 |\partial_V \hat{H}| E_0 \rangle}
{\langle E_0 | E_0 \rangle}
=C_V,
\end{eqnarray}
where 
\begin{eqnarray}
\label{thermo13}
C_V = \frac{\langle E_0 |\hat{\pi}_V| E_0 \rangle}
{\langle E_0 | E_0 \rangle}
=\sum_{\bm{n}\in \mathbb{F}: V(\bm{n})=V}
\frac{|\langle \bm{n} | E_0 \rangle|^2}{\langle E_0 | E_0 \rangle}
\end{eqnarray}
is the ground state expectation of the projector $\hat{\pi}_V$.
On the other hand, in the limit $M\to\infty$ 
we have $\partial_V E_0=\partial_v e_0$ so that 
by derivating Eq. (\ref{thermo1}) with respect to a generic $v$, 
we get, in the notation in which the potential levels are ordered,
\begin{equation}
\label{thermo14}
C_k= \frac{\partial e_0}{\partial v_{k}} =
\frac{p_k}{(v_k-e_0)^2} 
\left[
\sum_{k'=1}^{m_\mathscr{V}} \frac{p_{k'}}{(v_{k'}-e_0)^2} 
\right]^{-1},
\quad k=1,\ldots,m_\mathscr{V}.
\end{equation}
Equation (\ref{thermo14}) holds at any point of the 
$2(m_\mathscr{V}-1)$-dimensional space 
$(p_2,v_2-v_1,\ldots,p_{m_\mathscr{V}},v_{m_\mathscr{V}}-v_1)$.
If $p_1=0$, from Eqs. (\ref{thermo9}-\ref{thermo9b}) 
and (\ref{thermo14}) we get 
\begin{eqnarray}
C_1 =
\cases{
1, & $W \leq 1$,\\ 
0, & $W > 1$,\\
}
\end{eqnarray}
whereas for $k=2,\ldots,m_{\mathscr{V}}$
\begin{eqnarray}
C_k =
\cases{
0, & $W \leq 1$,\\
\frac{p_k}{(v_k-e_0)^2} 
\left[ \sum_{k'=2}^{m_\mathscr{V}} 
\frac{p_{k'}}{(v_{k'}-e_0)^2} \right]^{-1}, & $W > 1$.\\ 
}
\end{eqnarray}

\begin{figure}[t]
\centering
\includegraphics[width=0.7\columnwidth,clip]{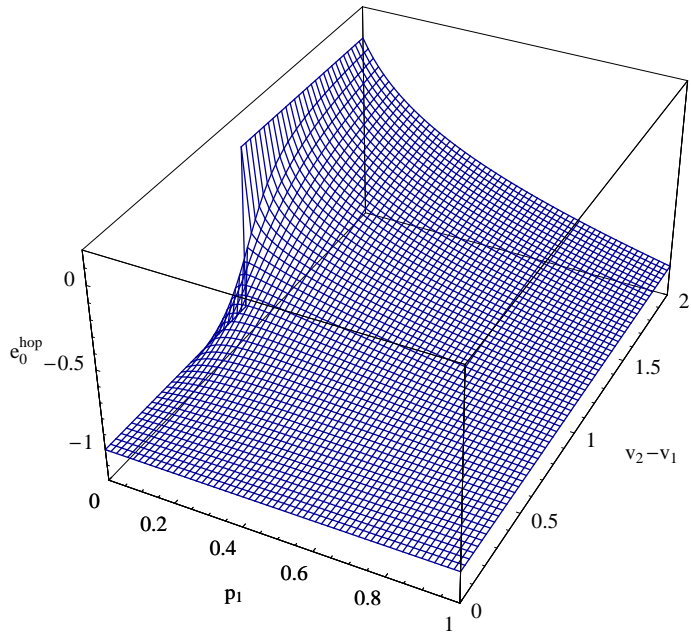}
\caption{ 
Thermodynamic limit of a system with two potential levels: 
asymptotic rescaled hopping energy $e_0^\mathrm{hop}$
as a function of $p_1$ and $v_2-v_1$.} 
\label{multinomial.fig4}
\end{figure}

We deduce that in the thermodynamic limit  
we have the following behavior of the ground state $|E_0\rangle$.
In the region $W>1$, $|E_0\rangle$ turns out to be an analytic 
function of its arguments 
$p_k,v_k-v_1$, $k=2,\ldots,m_\mathscr{V}$,
while in the region $W \leq 1$ it collapses into
the subspace spanned by the configurations with minimum potential value $V_1$.
The susceptibilities $\partial_{(v_k-v_1)}C_k$ diverge on the critical 
surface $\mathscr{S}$ determined by the condition $W=1$.
Finally, it is simple to check that the asymptotic rescaled hopping energy
in the ground state,
$e_0^\mathrm{hop} = \lim_{M\to\infty} 
\langle E_0(M) |\hat{K}| E_0(M) \rangle/|E_0^{(0)}(M)|$,
is given by 
\begin{equation}
\label{hop}
e_0^\mathrm{hop}= 
-\left[ \sum_{k=1}^{m_\mathscr{V}} 
\frac{p_{k}}{(v_{k}-e_0)^2} \right]^{-1}.
\end{equation}
Therefore, the phase for $W\leq 1$ is a frozen phase with 
$e_0^\mathrm{hop}=0$, 
whereas a non vanishing hopping energy is obtained for $W>1$.
In Fig. \ref{multinomial.fig4} we show the behavior of
$e_0^\mathrm{hop}(p_1,v_2-v_1)$
in the case with two potential levels described above.
For $p_1=0$, it is evident the discontinuity of $e_0^\mathrm{hop}$
at the critical point $W=1$, \textit{i.e.} $v_2-v_1=1$.

\section{Many-body lattice models: random potential with discrete spectrum}
\label{many-body.discrete}

Hereinafter we will focus our analysis 
on the thermodynamic limit of random potential systems. 
Equation (\ref{thermo1}) states that for these systems the asymptotic 
rescaled energy $e_0$ is universal, independently of the nature of the 
hopping operator, provided that $\lim_{M\to\infty}|E_0^{(0)}(M)|=\infty$.
In this Section we present the results of numerical simulations 
which help to quantify the rapidity with which this universality 
is approached by systems with different complexity.

\begin{figure}[t]
\centering
\includegraphics[width=0.8\columnwidth,clip]{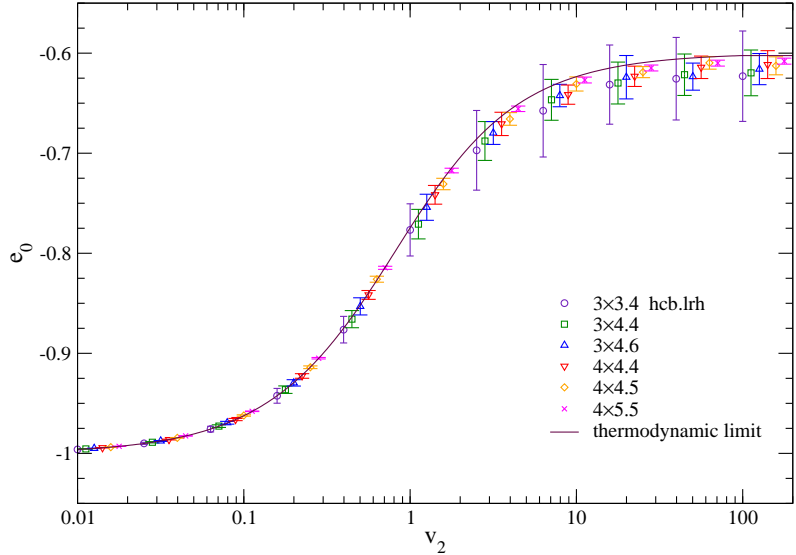}
\caption{Universality of the thermodynamic limit of random potential systems: 
rescaled ground-state energy $e_0$ as a function of the 
rescaled potential level $v_2$ for systems
of spinless hard-core bosons with long range hopping. 
The random potential has two levels: $V_1=0$ with probability $p_1=0.6$ and
$V_2=v_2 |E_0^{(0)}|$ with probability $p_2=0.4$.
The solid line is the universal thermodynamic limit predicted by Eq. 
(\ref{thermo1})
whereas the dots corresponds to the numerical results found for the 
finite-size systems indicated in the legend.
Here, $m$$\times$$n.N$ connotes a system of $N$ particles in a 
$m \times n$ lattice. 
The error bars represent the standard deviation of the stochastic 
variable $e_0$ as evaluated from an ensemble of 100 exact diagonalizations 
of the Hamiltonian matrix with the specified random potential.}
\label{multinomial.fig5}
\end{figure}

\begin{figure}[t]
\centering
\includegraphics[width=0.8\columnwidth,clip]{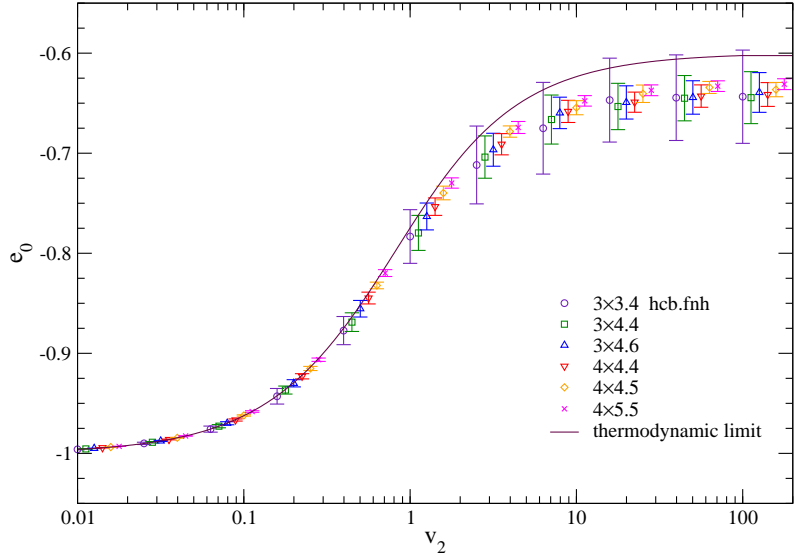}
\caption{As in Fig. \ref{multinomial.fig5} in the case of systems 
of spinless hard-core bosons with first neighbor hopping.}
\label{multinomial.fig6}
\end{figure}

\begin{figure}[t]
\centering
\includegraphics[width=0.8\columnwidth,clip]{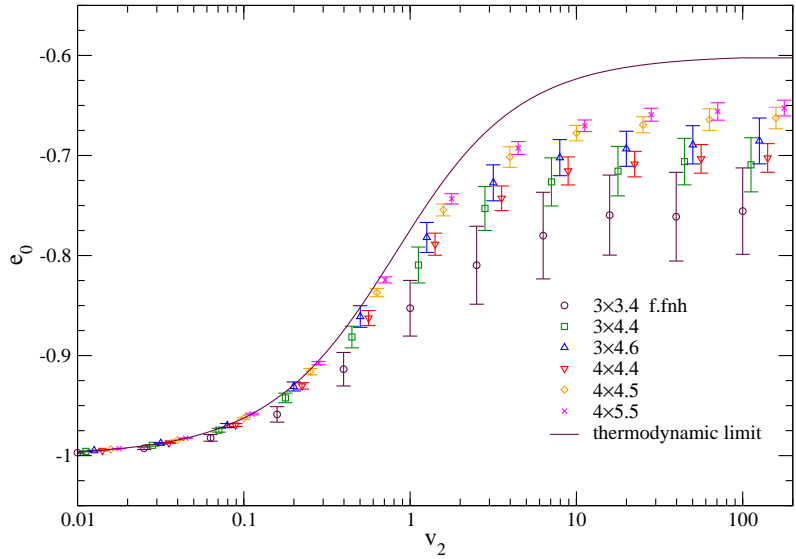}
\caption{As in Fig. \ref{multinomial.fig5} in the case of systems 
of spinless fermions with first neighbor hopping.}
\label{multinomial.fig7}
\end{figure}

To begin, we investigate quantum particles moving in two-dimensional 
lattices and interacting via a potential with two random levels. 
Three cases have been considered: 
spinless hard-core bosons with long range hopping, 
spinless hard-core bosons and spinless fermions with first neighbor hopping. 
The results are displayed in Figs. \ref{multinomial.fig5}, 
\ref{multinomial.fig6} and \ref{multinomial.fig7}, respectively.
For these systems, in Table \ref{table2D} we report the noninteracting 
ground-state energy $E_0^{(0)}$ and the average number of active links 
\begin{equation}
\label{Ast}
  A_\mathrm{av} = \frac{1}{M} \sum_{\bm{n}\in\mathbb{F}} A(\bm{n})
\end{equation}
as a function of the dimension $M$ of the Fock space.
In the case of fermions, due to sign cancellations, the effective average 
number of active links which is responsible of the potential 
decorrelation is the difference 
\begin{equation}
\label{Aeff}	
  A_\mathrm{av}= \frac{1}{M} \sum_{\bm{n}\in\mathbb{F}} 
\left( A_+(\bm{n}) - A_-(\bm{n}) \right)
\end{equation}
between the active links $A_\pm(\bm{n})$ associated with positive or negative
jumps, respectively.

\begin{figure}[t]
\centering
\includegraphics[width=0.8\columnwidth,clip]{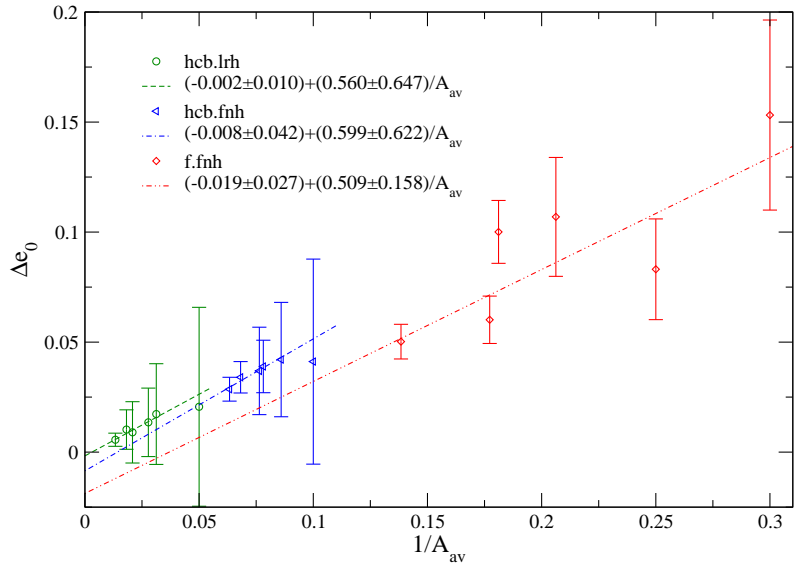}
\caption{Difference $\Delta e_0$ between the rescaled ground-state energy 
$e_0$ predicted in the thermodynamic limit for $v_2=100$ 
by the multinomial formula (\ref{thermo1}) and that obtained 
in the finite-size systems analyzed in
Fig.~\ref{multinomial.fig5} (hcb.lrh $\circ$), 
Fig.~\ref{multinomial.fig6} (hcb.fnh $\triangleleft$) and 
Fig.~\ref{multinomial.fig7} (f.fnh $\diamond$)
as a function of the inverse of the average number of active links 
$A_\mathrm{av}$.
The lines are a weighted linear fit with errors on $\Delta e_0$ assumed    
equal to the standard deviations of $e_0$ for the corresponding 
finite-size systems.} 
\label{multinomial.fig8}
\end{figure}
\begin{table}
  \caption{\label{table2D}For different systems $m$$\times$$n.N$
of $N$ particles in a two-dimensional $m$$\times$$n$ lattice 
with periodic boundary conditions  
we report the number of states $M$, the noninteracting ground-state energy 
$E_0^{(0)}$ and the average number of active links $A_\mathrm{av}$ in
the case of spinless hard-core bosons with long-range (hcb.lrh) 
and first-neighbor hopping (hcb.fnh) and 
spinless fermions with first-neighbor hopping (f.fnh).}
  \lineup
  \begin{indented}
  \item[]\begin{tabular}{@{}llllllll}
    \br
    & & \centre{2}{hcb.lrh} & \centre{2}{hcb.fnh} & \centre{2}{f.fnh}\\
    \ns\ns
    & &\crule{2}&\crule{2}&\crule{2}\\
    system & \0\0\0$M$ & $E_0^{(0)}$ & $A_\mathrm{av}$ & \0$E_0^{(0)}$ &$A_\mathrm{av}$ & \0$E_0^{(0)}$ &$A_\mathrm{av}$\\
    \mr
    3$\times$3.4 & \0\0126 & -20 & 20 & -10.34699 & 10       &      \0-7 & 3.33333\\
    3$\times$4.4 & \0\0495 & -32 & 32 & -12.10548 & 11.63636 &      \0-9 & 4.84848\\
    3$\times$4.6 & \0\0924 & -36 & 36 & -13.71426 & 13.09090 &       -10 &      4 \\
    4$\times$4.4 &  \01820 & -48 & 48 & -13.25169 & 12.8     &       -10 & 5.52087\\
    4$\times$4.5 &  \04368 & -55 & 55 & -15.30041 & 14.66666 &       -12 & 5.64102\\
    4$\times$5.5 &   15504 & -75 & 75 & -16.44602 & 15.78947 & -13.23607 & 7.22394\\
    \br
  \end{tabular}
  \end{indented}
\end{table}

The case of hard-core bosons with long range hopping presents the closest
analogy with a uniformly fully connected model.
The number of active links $A(M)$, although being much smaller than the
number of states $M$ and diverging more slowly than $M$ for $M\to\infty$, 
is a constant which does not depend on the configuration $\bm{n}$. 
In particular, we have that $A(M)=A_\mathrm{av}(M)=-E_0^{(0)}(M)$ is given by
the simple formula $(\textrm{number of sites}-\textrm{number of particles})
\times\textrm{number of particles}$. 
The numerical results, shown in Fig. \ref{multinomial.fig5} by dots 
with error bars, have been obtained by exact diagonalizations 
of the Hamiltonian matrix for several realizations of the random potential.
The error bars represent the standard deviation of the stochastic
variable $E_0/|E_0^{(0)}|$ evaluated in this way.
The agreement with the universal curve for $e_0$, obtained from Eq. 
(\ref{thermo1}) as a function of the second rescaled potential level $v_2$,
worsen by increasing the value of $v_2$. 
However, we have a clear tendency toward the universal result by choosing 
systems closer and closer to the thermodynamic limit. 

For systems with first neighbor hopping, we have a much greater complexity 
with respect to a uniformly fully connected model. 
The number of active links is a function of the configuration $\bm{n}$
and, in the case of fermions, a sign problem is also present. 
Therefore, we expect a convergence to the universal thermodynamic limit
slower than that obtained with long range hopping.
This is confirmed by the results shown in Figs. \ref{multinomial.fig6} and 
\ref{multinomial.fig7}.
Whereas the trend toward the universal behavior is clear in both cases, 
for fermions the convergence is indubitably slower than for hard-core bosons. 

Note that for practical reasons, essentially the finite amount 
of memory (4 Gb) of the computer used to perform the numerical 
diagonalizations,
the different systems considered in Figs. \ref{multinomial.fig5}, 
\ref{multinomial.fig6} and \ref{multinomial.fig7} do not have exactly the same 
density, $\textrm{number of particles}/\textrm{number of lattice sites}$,
as it would be convenient in looking at the thermodynamic limit of a 
lattice particle system. 
In fact, we have chosen a set of systems in which 
$M$ and $|E_0^{(0)}|$, \textit{i.e.} $A_\mathrm{av}$, 
possibly both increase compatibly with the 
condition $M \leq M_\mathrm{max}$, where $M_\mathrm{max}$ is the 
size of the largest diagonalizable Hamiltonian matrix. 
At constant density this set would contain only two or three elements,
so we decided to allow for density fluctuations.
These fluctuations may reflect
a non monotonous approaching to the thermodynamic limit.
Such a behavior is quite evident in the fermion case of 
Fig. \ref{multinomial.fig7}
where the system 3$\times$4.6 with $M=924$ and $A_\mathrm{av}=4$
is closer to the asymptotic universal curve than the next system 
4$\times$4.4 which has $M=1820$ and $A_\mathrm{av}=5.52087$.

In Fig. \ref{multinomial.fig8} we show a quantitative test 
for the convergence of data in Figs.~\ref{multinomial.fig5}, 
\ref{multinomial.fig6} and \ref{multinomial.fig7}.
The difference $\Delta e_0$ between the rescaled ground-state energy
of the finite-size systems and the thermodynamic value predicted by the 
multinomial formula (\ref{thermo1}) is reported, for $v_2=100$,
as a function of $A_\mathrm{av}^{-1}$.
In all cases, the data are fitted by straight lines which, 
compatibly with the associated errors, give $\Delta e_0 \to 0$ for
$A_\mathrm{av}^{-1}\to 0$. 
This behavior corresponds to the scenario depicted in Section \ref{rps}: 
as the average number of active links $A_\mathrm{av}$ diverges, 
the multinomial formula (\ref{thermo1}) becomes exact
with a residual error proportional to $A_\mathrm{av}^{-1}$. 
This law also explains the progressively larger errors observed,
at a fixed system size, 
passing from hard-core bosons with long range hopping to
hard-core bosons with first neighbor hopping to
fermions with first neighbor hopping, 
see Eqs. (\ref{Ast}) and (\ref{Aeff}).

The models considered above, even if share some realistic features
with systems of interest in physics, 
have been studied in connection with a toy random potential defined 
by only two levels $V_1$ and $V_2$ with assigned probabilities 
$p_1$ and $p_2$.
For these models, in order to obtain the phase transition
in the ground state we must impose the additional condition $p_1\to 0$
after the thermodynamic limit has been taken.
In the remaining part of this Section, we discuss a more realistic  
random potential which has the following characteristics:
\textit{i)} the number of levels increases by increasing the number 
of states $M$, and 
\textit{ii)} the probability associated with the lowest level vanishes for
$M\to \infty$.

Consider a lattice $\Lambda$ with some ordering of the lattice points 
$1,2,\dots,n_{|\Lambda|}$ occupied by $N$ quantum particles.
As above, we assume that the particles are fermions or hard-core bosons
so that the occupation numbers can be $n_i=0,1$, $i=1,2,\dots,|\Lambda|$. 
To each of the $M=|\Lambda|!/(N! (|\Lambda|-N)!)$ Fock states
$\bm{n}=(n_1,n_2,\dots,n_{|\Lambda|})$ of the system
we associate the potential
\begin{equation}
  \label{impurity.potential}
  V(\bm{n}) = \gamma \sum_{i=1}^{|\Lambda|} n_i r_i(\bm{n}), 
\end{equation}
where $r_i(\bm{n})$ are a set of $|\Lambda| M$ 
independent random variables assuming the values $0,1$.
Let $p$ be the probability that $r_i(\bm{n})=1$.
Since the Fock states are normalized by the condition
$\sum_{i=1}^{|\Lambda|} n_i = N$,
we have $N+1$ different potential levels $V_k=\gamma k$, $k=0,1,\dots,N$
with associated probabilities 
\begin{eqnarray}
  p_k &=& \frac{N!}{k! (N-k)!} ~p^k (1-p)^{N-k}.
\end{eqnarray}
Note that $p_0$, the probability of the lowest potential level $V_0=0$, 
vanishes for $N\to\infty$ as far as $p>0$.

Before going further, we want to spend a few comments about the 
introduced potential (\ref{impurity.potential}).
It represents the on site random coupling of a set of $|\Lambda|$ 
impurities with the particles of the system. 
The impurities, spins to be concrete, may be in one of two available 
states with probabilities $p$ and $1-p$. 
Impurity-impurity interactions are neglected. 
Moreover, the potential (\ref{impurity.potential}) treats the impurities
itself as classical spins, \textit{i.e.} it does not take into account 
the correlations of the state of each single impurity 
with itself in the presence of different dispositions of the particles 
in the lattice (Fock states).
The latter decoupling could be the effective result of an impurity 
dynamics fast on the time scale of the evolution of the 
particle system or of the presence of an external random field.

In the thermodynamic limit $|\Lambda|,N\to \infty$ with $N/|\Lambda|$ constant,
the rescaled ground-state energy $e_0= E_0/|E_0^{(0)}|$ is determined by
Eq. (\ref{thermo1}).
In this limit the sum over the discrete rescaled levels
$v_k=V_k/|E_0^{(0)}|$ can be approximated by an integral over $k$
so that Eq. (\ref{thermo1}) reads 
\begin{equation}
  \label{thermo_impurity}
  \int_0^N \frac{p_k}{v_k - e_0} dk =1, 
  \qquad e_0\leq 0.
\end{equation}
Since Eq. (\ref{impurity.potential}) is a linear combination of the 
independent random variables $r_i(\bm{n})$, in the thermodynamic limit
the distribution of the rescaled potential values, $v_k$, 
becomes infinitely peaked around the mean value $\bar{v}=v_{Np}$,
\begin{eqnarray}
  p_k \to \delta(k-Np),
\end{eqnarray}
which inserted into Eq. (\ref{thermo_impurity}) gives 
\begin{equation}
  \frac{1}{v_{Np} - e_0}=1, 
  \qquad e_0\leq 0.
\end{equation}
In conclusion, the solution for the rescaled ground-state energy is
\begin{eqnarray}
\label{e0.impurity}
e_0 =
\cases{
-1+\gamma Np/|E_0^{(0)}|, & $\gamma Np/|E_0^{(0)}| < 1$,\\ 
0, & $\gamma Np/|E_0^{(0)}| \geq 1$.\\
}
\end{eqnarray}

\begin{figure}[t]
\centering
\includegraphics[width=0.8\columnwidth,clip]{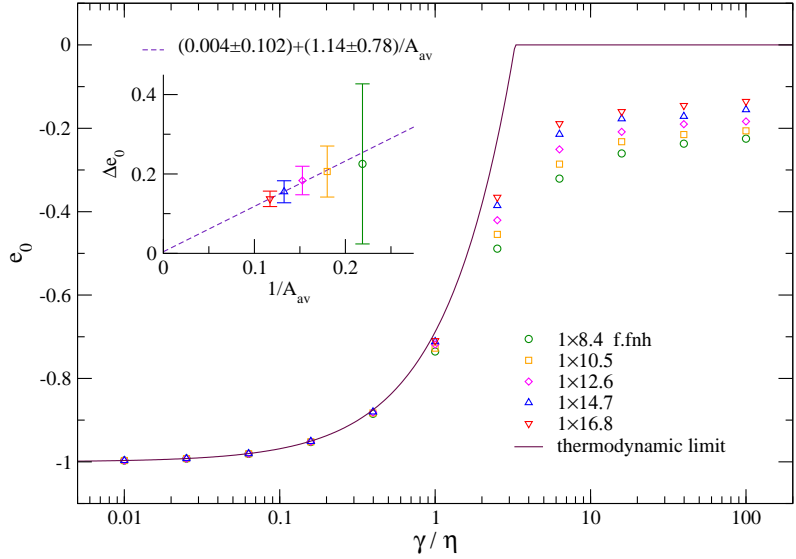}
\caption{
Universality of the thermodynamic limit of random potential systems: 
rescaled ground-state energy $e_0$ as a function of the 
rescaled potential strength $\gamma/\eta$ for systems,
1$\times$$n.N$,
of $N$ spinless fermions with first neighbor hopping
in a one-dimensional lattice with $n$ sites.
The potential is the impurity potential of Eq. (\ref{impurity.potential})
with $p=0.4$.  
The solid line is the universal thermodynamic limit predicted by Eq. 
(\ref{e0.impurity.f.fnh}) 
whereas the dots corresponds to the numerical results found for the 
finite-size systems indicated in the legend.
The inset shows a linear fit of $\Delta e_0$ \textit{vs.} 
$A_\mathrm{av}^{-1}$ as displayed in Fig.~\ref{multinomial.fig8}.}
\label{multinomial.fig9}
\end{figure}

Whether or not the solution given above 
represents a non trivial rescaled ground-state energy depends on
how fast the potential levels diverge in the thermodynamic limit 
with respect to the noninteracting ground-state energy.
For a system of fermions with first neighbor hopping 
$E_0^{(0)}$ can be evaluated analytically 
by Fourier transformation. 
In one dimension with periodic boundary conditions, 
at half filling and neglecting spin, 
for $N \to \infty$ we have  
$E_0^{(0)} \simeq -4\pi^{-1} N \eta$, where $\eta$ is the value of the
hopping coefficient between any two first neighbors. 
Equation (\ref{e0.impurity}) then becomes
\begin{eqnarray}
\label{e0.impurity.f.fnh}
e_0 =
\cases{
-1+ \pi \gamma p/ 4 \eta, & $\pi \gamma p < 4\eta$,\\ 
0, & $\pi\gamma p \geq 4\eta$.\\
}
\end{eqnarray}
A comparison of $e_0$ predicted by the above expression for $p=0.4$ 
with the values obtained from numerical simulations of finite-size systems 
is shown in Fig.~\ref{multinomial.fig9}.
The main parameters of the simulated systems are given in Table~\ref{table1D}.
The numerical results are consistent with a phase transition occurring 
in the thermodynamic limit at $\gamma/\eta=4/\pi p \simeq 3.18$.

\begin{figure}[t]
\centering
\includegraphics[width=0.8\columnwidth,clip]{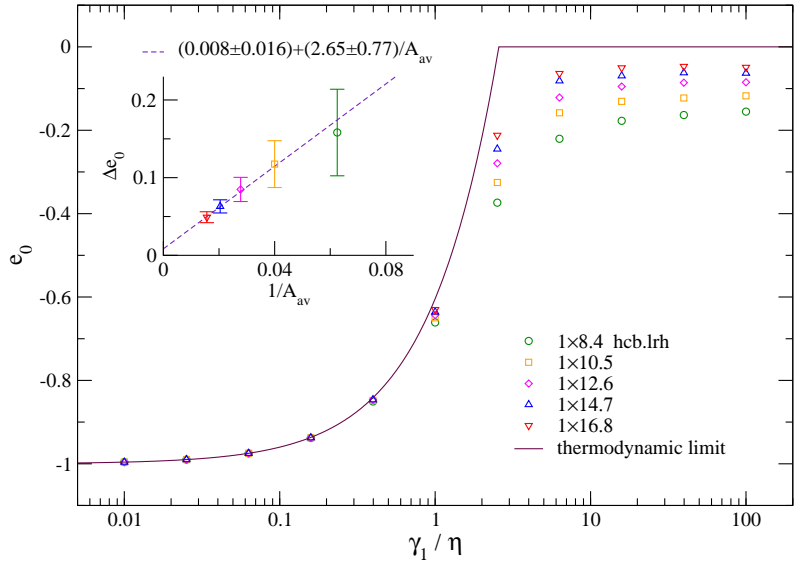}
\caption{
As in Fig. \ref{multinomial.fig9} in the case of spinless hard-core bosons 
with long range hopping. The potential is the impurity potential of 
Eq. (\ref{impurity.potential}) with $p=0.4$ and $\gamma=\gamma_1 N$, 
with $\gamma_1$ constant.}
\label{multinomial.fig10}
\end{figure}
\begin{table}
  \caption{\label{table1D}As in table \ref{table2D} in the case of 
one-dimensional systems $1$$\times$$n.N$.}
  \lineup
  \begin{indented}
  \item[]\begin{tabular}{@{}llllllll}
    \br
    & & \centre{2}{hcb.lrh} & \centre{2}{hcb.fnh} & \centre{2}{f.fnh}\\
    \ns\ns
    & &\crule{2}&\crule{2}&\crule{2}\\
    system & \0\0\0$M$ & $E_0^{(0)}$ & $A_\mathrm{av}$ & \0$E_0^{(0)}$ &$A_\mathrm{av}$ & \0$E_0^{(0)}$ &$A_\mathrm{av}$\\
    \mr
    1$\times$8.4 & \0\0\070 & -16 & 16 & \0-5.22625 & 4.57142 & \0-4.82842 & 3.42857\\
    1$\times$10.5 & \0\0252 & -25 & 25 & \0-6.47213 & 5.55555 & \0-6.47213 & 5.55555\\
    1$\times$12.6 & \0\0924 & -36 & 36 & \0-7.72741 & 6.54545 & \0-7.46410 & 5.45454\\
    1$\times$14.7 &  \03432 & -49 & 49 & \0-8.98792 & 7.53846 & \0-8.98791 & 7.53846\\
    1$\times$16.8 &   12870 & -64 & 64 &  -10.25166 & 8.53333 &  -10.05467 & 7.46666\\
    \br
\lineup
  \end{tabular}
  \end{indented}
\end{table}

The value of the noninteracting ground-state energy can be given analytically
also for spinless hard-core bosons with long range hopping.
At half filling with periodic boundary conditions we have 
$E_0^{(0)} = -N^2 \eta$, where $\eta$ is the value of the
hopping coefficient between any two sites.
In this case, for $\gamma$ constant the potential levels diverge for
$N\to\infty$ slower than $E_0^{(0)}$ and Eq. (\ref{e0.impurity}) 
always gives $e_0=-1$. 
A non trivial result is obtained assuming $\gamma=\gamma_1 N$, 
with $\gamma_1$ constant,
\begin{eqnarray}
\label{e0.impurity.hcb.lrh}
e_0 =
\cases{
-1+ \gamma_1 p/\eta, & $\gamma_1 p < \eta$,\\ 
0, & $\gamma_1 p \geq \eta$.\\
}
\end{eqnarray}
In Fig.~\ref{multinomial.fig10} we show the behavior of
Eq. (\ref{e0.impurity.hcb.lrh}) for $p=0.4$ in comparison 
with numerical simulations for finite-size systems. 
Again, there is a consistent matching with the singularity 
displayed by $e_0$ at $\gamma_1/\eta=1/p=2.5$.
As discussed in Section \ref{thermolimit}, 
in the present case the approach to the thermodynamic behavior is faster
(smaller differences $\Delta e_0$) with respect to the case 
of spinless fermions with first neighbor hopping shown in 
Fig.~\ref{multinomial.fig9}.

Note that, for clarity, we have plotted in Figs.~\ref{multinomial.fig9} 
and \ref{multinomial.fig10} different curves for $e_0$.
However, as it is evident by Eqs. (\ref{e0.impurity.f.fnh})
and (\ref{e0.impurity.hcb.lrh}), the behavior of the
rescaled ground-state energy is universal once it is expressed 
in terms of the rescaled potential levels $V_k/|E_0^{(0)}|$, 
provided, of course, that the same probabilities $p_k$ are used.

\section{Many-body lattice models: random potential with continuous spectrum}
\label{many-body.continuous}

The results obtained in Section \ref{thermolimit} are readily 
extended to the case of a random potential with continuous spectrum.
In the thermodynamic limit, for a continuous distribution of rescaled levels 
$v= \lim_{M\to\infty} V(M)/|E_0^{(0)}(M)|$ 
described by the density $p(v)=\lim_{M\to\infty} p(V(M))$,
the equation determining the rescaled energy 
$e_0 = \lim_{M\to\infty} E_0(M)/|E_0^{(0)}(M)|$ reads
\begin{eqnarray}
\label{cthermo1}
\int
 \frac{p(v)}{v-e_0}dv=1,
\qquad e_0\leq v_{\mathrm{min}}.
\end{eqnarray}
The only point which we have to pay attention to concerns the definition 
of the lowest potential level. 
In order to avoid ambiguities, we define $v_\mathrm{min}$ as the value 
of $v$ such that
\begin{equation}
\label{cvmin}
\begin{array}{c}
p(v_\mathrm{min}+\delta)>0 \qquad \mbox{and} \qquad
p(v_\mathrm{min}-\delta)=0,
\end{array}
\end{equation}
with $\delta>0$ arbitrarily small.
Note that definition (\ref{cvmin}) does not include 
the case $v_\mathrm{min}=-\infty$, which occurs, for example, 
if the density $p(v)$ is a Gaussian.
However, since for $v_\mathrm{min}=-\infty$ we can only have the 
trivial result $e_0=v_\mathrm{min}=-\infty$, 
in the following we will assume $v_\mathrm{min}$ finite.

In order to recover the analogous results of Section \ref{thermolimit}, 
it is useful to define
\begin{eqnarray}
\label{cF}
F(x) = \int \frac{p(v)}{v-x}dv
\end{eqnarray}
and
\begin{eqnarray}
\label{cthermoW}
W = F(v_{\mathrm{min}}) = \int\frac{p(v)}{v-v_{\mathrm{min}}}dv.
\end{eqnarray}
Observe that $F(x)$ is a positive monotonously increasing function 
for $x \in (-\infty,v_\mathrm{min}]$ with 
\begin{eqnarray}
\label{csup}
-\infty < F(x) \leq W=\sup F 
\qquad x \in (-\infty,v_\mathrm{min}].
\end{eqnarray}
We have that $e_0$ is solution of Eq. (\ref{cthermo1})
if $F(e_0)=1$ and $e_0\leq v_{\mathrm{min}}$. 
If $p(v_{\mathrm{min}})>0$, and therefore $W=\infty$,  
Eq. (\ref{csup}) shows that the solution $e_0$ exists, 
it is unique and smooth, $\textit{i.e.}$
without singularities with respect to the parameters of the
density $p(v)$. 
If $p(v_\mathrm{min})=0$ then, since $\int p(v)dv=1$, we have $W<\infty$.
For $1 \leq W < \infty$, we still have a unique smooth solution $e_0$
of Eq. (\ref{cthermo1}), with $e_0 < v_{\mathrm{min}}$ for $W>1$
and $e_0 = v_{\mathrm{min}}$ for $W=1$. 
For $W < 1$, Eq. (\ref{cthermo1}) has no solution and,
as discussed in the discrete case, we have to consider for $e_0$ 
its analytic continuation for $p(v_\mathrm{min})\to0$.
More explicitly, we first solve Eq. (\ref{cthermo1}) with a modified
density $\tilde{p}(v)$ having the same support of $p(v)$ but with 
$\tilde{p}(v_\mathrm{min}) >0$ and then
evaluate $e_0$ as the limit for $\tilde{p} \to p$ of the solution so found.
Clearly, this analytic continuation cannot exceed $v_{\mathrm{min}}$
and it is easy to demonstrate that we have $e_0=v_\mathrm{min}$.  
In fact, if it were $e_0<v_\mathrm{min}$, 
there would be no singularity in the integrand of Eq. (\ref{cthermo1})
and the analytic continuation of $e_0$ for $\tilde{p} \to p$ 
would coincide with the solution of Eq. (\ref{cthermo1}) with density $p$.
As a consequence, we should have $W>F(e_0)=1$ in contradiction with 
the hypothesis $W<1$.
In conclusion, for $p(v_{\mathrm{min}})=0$ the characterization of 
the ground state in terms of $W$ is the same as in the discrete case
\begin{eqnarray}
\label{cthermo9}
W > 1 \Leftrightarrow  e_0<v_{\mathrm{min}}, \\
\label{cthermo9b}
W\leq1 \Leftrightarrow  e_0=v_{\mathrm{min}}.
\end{eqnarray}

Even when the potential levels $V$ are continuous random variables, 
the definition (\ref{thermo10}) and the relations
(\ref{thermo12}) and (\ref{thermo13}) remain unchanged, and
by performing functional derivatives of the rescaled ground-state energy
$e_0$ with respect to $v$, we have
\begin{equation}
\label{cthermo14}
C(v)= \frac{\delta e_0}{\delta v} =
\frac{p(v)}{(v-e_0)^2} 
\left[
\int\frac{p(v')}{(v'-e_0)^2}dv' 
\right]^{-1}.
\end{equation}
Equation (\ref{cthermo14}) holds for any density $p(v)$. 
Again, if $p(v_\mathrm{min})=0$, from Eqs. (\ref{cthermo9}-\ref{cthermo9b}) 
and (\ref{cthermo14}) we get 
\begin{eqnarray}
\label{cvminc}
C(v_\mathrm{min}) =
\cases{
1, & $W \leq 1$,\\ 
0, & $W > 1$,\\
}
\end{eqnarray}
whereas for $v>v_\mathrm{min}$
\begin{eqnarray}
\label{cvc}
C(v) =
\cases{
0, & $W \leq 1$,\\
\frac{p(v)}{(v-e_0)^2} 
\left[
\int\frac{p(v')}{(v'-e_0)^2}dv' 
\right]^{-1}, & $W > 1$.\\ 
}
\end{eqnarray}

Finally, it is simple to extend Eq. (\ref{hop}) for the rescaled 
hopping energy in the ground state,
\begin{equation}
\label{chop}
e_0^\mathrm{hop}=- 
\left[
\int\frac{p(v)}{(v-e_0)^2}dv
\right]^{-1}.
\end{equation}
Again, if $p(v_\mathrm{min})=0$, we see that
whereas a non vanishing hopping energy is obtained for $W>1$,
the phase for $W\leq 1$ is a frozen phase with $e_0^\mathrm{hop}=0$.

\begin{figure}
\centering
\includegraphics[width=0.8\columnwidth,clip]{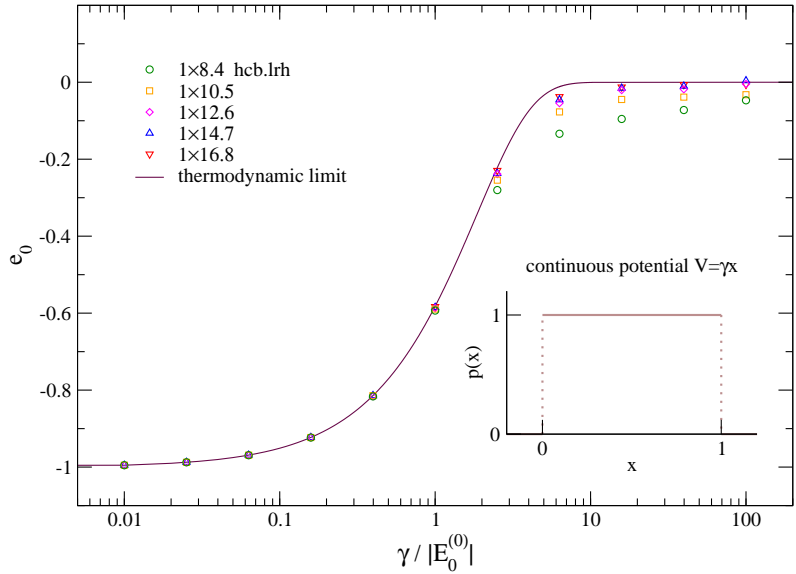}
\caption{
Universality of the thermodynamic limit of random potential systems: 
rescaled ground-state energy $e_0$ as a function of the 
rescaled potential strength $\gamma/|E_0^{(0)}|$ for systems,
1$\times$$n.N$,
of $N$ spinless hard-core bosons with long range hopping
in a one-dimensional lattice with $n$ sites.
The potential randomly associated with the states is $V=\gamma x$,
where $x$ is a random variable in the interval $[0,1]$ 
with normalized constant density $p(x)=1$.
The solid line is the universal thermodynamic limit predicted by 
Eq. (\ref{cthermo1}) whereas the dots are the numerical results for 
the finite-size systems indicated in the legend.} 
\label{multinomial.fig11}
\end{figure}

\begin{figure}
\centering
\includegraphics[width=0.8\columnwidth,clip]{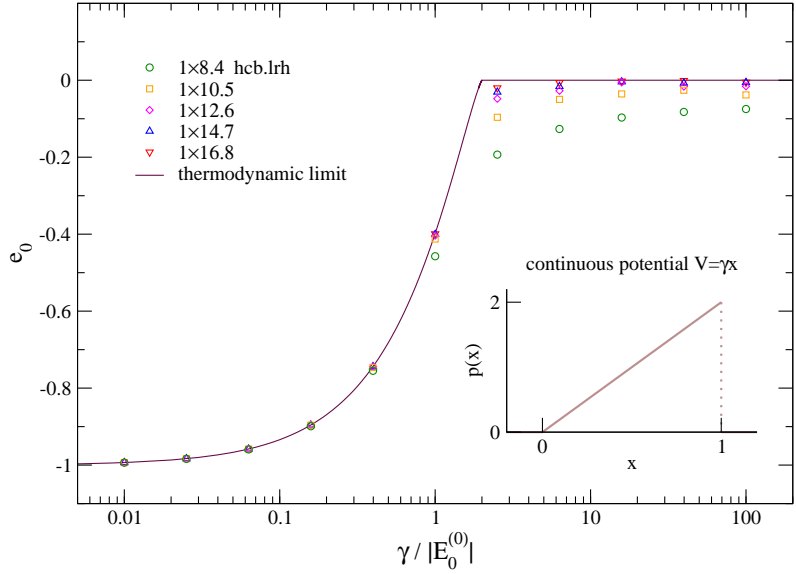}
\caption{
As in Fig. \ref{multinomial.fig11} in the case of the potential 
$V=\gamma x$, where $x$ is a random variable in the interval $[0,1]$ 
with normalized linear density $p(x)=2x$.} 
\label{multinomial.fig12}
\end{figure}

We conclude this Section by discussing the results of numerical simulations 
on finite-size systems with a random potential having a continuous spectrum.
The systems considered are 
spinless hard-core bosons with long range hopping in a one-dimensional lattice.
Similar results, not shown here, are obtained for 
spinless hard-core bosons and spinless fermions with first neighbor hopping.
The relevant parameters are given in Table \ref{table1D}.
No relevant differences are observed in two-dimensional lattices.
In Figs. \ref{multinomial.fig11} and \ref{multinomial.fig12}
we illustrate the results for two distinct potentials.
In both cases we have $V=\gamma x$
where $x$ is a random variable in the interval $[0,1]$ 
with normalized constant density $p(x)=1$ in Fig. \ref{multinomial.fig11} 
and with normalized linear density $p(x)=2x$ 
in Fig. \ref{multinomial.fig12}. 
Equation (\ref{cthermo1}) predicts, in the thermodynamic limit,  
a smooth rescaled ground-state energy
$e_0=E_0/|E_0^{(0)}|$ for the potential in Fig. \ref{multinomial.fig11}
and a phase transition for the potential in Fig. \ref{multinomial.fig12}.
In the latter case, the singularity is localized by the condition $W=1$,
\textit{i.e.} at $\gamma/|E_0^{(0)}|=2$.
The results from the finite-size systems in 
Figs. \ref{multinomial.fig11} and \ref{multinomial.fig12}
clearly show convergence towards the predicted thermodynamic behavior.    
\begin{figure}
\centering
\includegraphics[width=0.8\columnwidth,clip]{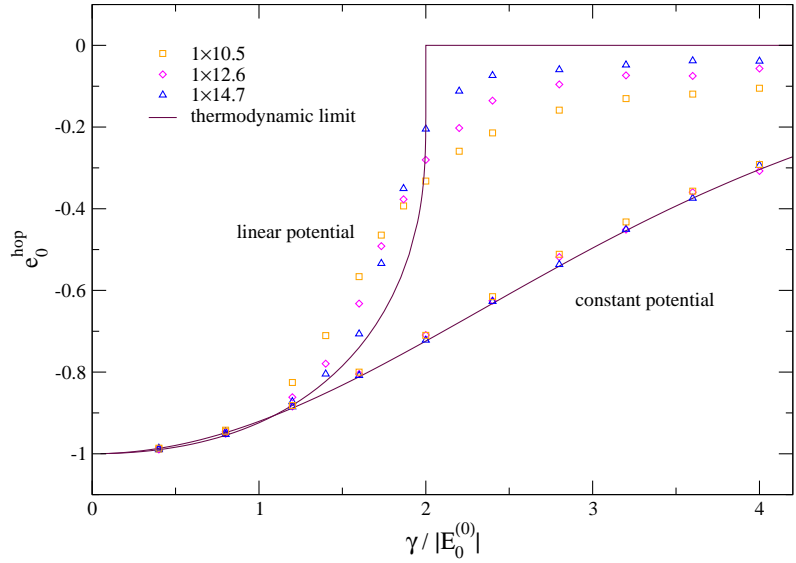}
\caption{
Rescaled hopping energy in the ground state $e_0^\mathrm{hop}$
as a function of the rescaled potential strength $\gamma/|E_0^{(0)}|$ 
for the systems with constant and linear random potential 
considered in Figs. \ref{multinomial.fig11} and \ref{multinomial.fig12}.
The solid lines are the thermodynamic values predicted by Eq. \ref{chop}
whereas the dots are the numerical results for the finite-size systems 
indicated in the legend.}
\label{multinomial.fig13}
\end{figure}

Finally, in Fig. \ref{multinomial.fig13} we show the
rescaled hopping energy in the ground state, $e_0^\mathrm{hop}$,
as a function of the rescaled potential strength for the two potentials
considered in Figs. \ref{multinomial.fig11} and \ref{multinomial.fig12}.
In the case of constant potential, we have a smooth
transition between the values $e_0^\mathrm{hop}=-1$ at $\gamma=0$ 
and $e_0^\mathrm{hop}\to 0$ for $\gamma \gg |E_0^{(0)}|$.
On the other hand, in the case of the linear potential there is a
discontinuity in the first derivative of $e_0^\mathrm{hop}$ as a function
of $\gamma$ at $\gamma/|E_0^{(0)}|=2$.
This behavior is consistent with that observed in the
simulations of finite-size systems.  

The random potential systems considered in this Section 
are a many-body generalization of the well-known Anderson models 
\cite{Anderson}, namely a single particle in a 
$d$-dimensional lattice in the presence of a disorder potential.
For the Anderson models with tight binding Hamiltonian the following 
rigorous results have been established \cite{FMSS}. 
For $d>1$, the eigenstates of the Hamiltonian  
undergo a localization phase transition for large disorder;
in one dimension, all states are localized for arbitrarily small non zero disorder.
It is clear that our analysis does not apply to the single-particle 
Anderson models, in which the number of active links, 
\textit{i.e.} the noninteracting ground-state energy $E_0^{(0)}$, 
remains constant in the limit $M\to\infty$. 
Note that in the single-particle case the Fock dimension $M$ 
coincides with the number of lattice points.
However, in the many-body case the phase transition predicted 
by Eq. (\ref{cthermo1}) has some analogies with the phase transition in the
Anderson models.
The transition of the ground state to the frozen phase characterized by
$e_0^\mathrm{hop} =0$ takes place, universally, for $W\leq 1$.
In \ref{AppendiceB} we show that this condition on the 
functional $W[p(\cdot)]$ is equivalent to a condition of large disorder, 
\textit{i.e.} large support of $p(v)$ or large potential strength.
Moreover, as predicted by Eqs. (\ref{cvminc}) and (\ref{cvc}) and 
also checked in the numerical simulations for finite-size systems,
in the frozen phase the ground state of the system $|E_0\rangle$ condensates
into a superposition of the Fock states which are eigenstates 
of the potential operator with eigenvalue $V=V_\mathrm{min}$.
This condensation, which is a sort of superlocalization in the Fock space, 
does not correspond, in general, to a localization
in the lattice space and takes place for any dimension $d$.

\section{Conclusions}
\label{conclusions}
We have revisited the probabilistic approach~\cite{OP1,OP2,OP4} 
to evaluate the ground state of general matrix Hamiltonian models
from a different point of view. 
The asymptotic probability density of the potential and hopping 
multiplicities which is the core of the approach is considered known 
and an exact equation for the ground-state energy $E_0$ is obtained 
in a closed form. 
For a class of systems, which includes the uniformly fully connected models 
and the random potential systems, we have that the above mentioned probability
density factorizes into separated potential and hopping contributions,
the potential part being of multinomial form.
This permits to write a very simple scalar equation 
which relates $E_0$ to the lowest eigenvalue of the Hamiltonian operator 
with zero potential $E_0^{(0)}$.
Such relation becomes exact in the thermodynamic limit
in which $E_0^{(0)}$ diverges.

The factorization of the probability density into separated potential 
and hopping contributions has a different origin in the uniformly fully 
connected models and in the random potential systems.
In the former case, due to the particular structure of the hopping 
operator which connects with equal probability any state to any other one, 
the potential values assumed by the system during its evolution are 
uncorrelated to the  number of allowed transitions, equal for any state.
In the latter case, the nature of the potential operator ensures that
the potential levels are associated randomly with the states of the system.
Correlations between potential and hopping values are, however,
reintroduced by the dynamics at a degree which depends on the structure
of the hopping operator. 
The correlations disappear if the number of allowed transitions
from any state diverges, 
\textit{i.e.} $A_{\mathrm{av}}\sim |E_0^{(0)}| \to\infty$.  
This explains why the relation that we have found between $E_0$ and $E_0^{(0)}$
becomes exact in the thermodynamic limit. 

In the thermodynamic limit, we find that the rescaled energy 
$e_0=E_0/|E_0^{(0)}|$ is a universal function of the potential levels, 
rescaled by $|E_0^{(0)}|$, and of the corresponding degeneracies, 
for the uniformly fully connected models, or probabilities, 
for the random potential systems.
In the case of the random potential systems this means that $e_0$
does not depend on the nature of the hopping operator. 
In general, if the degeneracy (probability) associated with the lowest 
potential level vanishes,
the ground state of the system undergoes a quantum phase transition between a 
normal phase and a frozen phase characterized by zero hopping energy.
The control parameter $W$ of the phase transition is related to the 
whole set of levels and corresponding degeneracies (probabilities) 
of the potential.
This parameter measures the inverse amount of the spread (disorder) 
introduced by the potential operator (random potential). 
In the frozen phase, corresponding to large spread (disorder), 
\textit{i.e.} small $W$, 
the ground state condensates into the subspace spanned by the states 
of the system associated with the lowest potential level.

We have performed numerical simulations on different finite-size 
random potential systems. 
In particular, we considered many-body lattice systems
in which the particles are fermions or hard-core bosons, 
with long range or first neighbor hopping,
in one or two dimensions.
The results found for the ground state are nicely consistent with the 
universal behavior predicted in the thermodynamic limit.

\section*{Acknowledgments}
This research was supported in part by Italian MIUR under 
PRIN 2004028108$\_$001
and by DYSONET under NEST/Pathfinder initiative FP6. 
M.O. acknowledges the Physics Department of the University of Aveiro 
for hospitality during the completion of this paper.

\appendix
\section{Proof of Equation (\ref{OMEGA})}
\setcounter{section}{1}
\label{Appendice}
Let us rewrite the probability density 
$\mathcal{P}_N(N\bm{\nu})$ in terms of its Fourier transform
$\widetilde{\mathcal{P}}_N(\bm{q})$
\begin{eqnarray}
\label{FT}
P_N(N\bm{\nu})=\int d\bm{q} 
~e^{\log[\widetilde{\mathcal{P}}_N(\bm{q})]-iN(\bm{q},\bm{\nu})}.
\end{eqnarray}
If we indicate with $\mediac{\nu_{\alpha_1}\ldots\nu_{\alpha_k}}$,
with $\alpha_1, \ldots, \alpha_k \in \mathscr{H}$,
a generic cumulant of order $k$, we have
\begin{eqnarray}
\label{LOGP}
\log {\widetilde{\mathcal{P}}}_N(\bm{q}) =
\log\media{e^{iN\left( \bm{\nu},\bm{q} \right)}}=
\sum_{k=1}^{\infty} \frac{N^k}{k!}
\mediac{\left( \bm{\nu},i\bm{q} \right)^k}.
\end{eqnarray}
For any given $N$, due to the inequalities 
$\media{\mu_{\alpha_1}\dots\mu_{\alpha_k}} \leq N^k$,
valid for any $k$, and due to the asymmetry 
$\mathcal{P}_N(\bm{\mu}) \neq \mathcal{P}_N(-\bm{\mu})$,
the series in Eq.~(\ref{LOGP}) converges 
for every $\bm{q} \in \mathbb{C}^m$  (see, for example~\cite{SHY}).

Let us introduce the rescaled cumulants 
of order $k$, in a compact notation $\bm{\Sigma}^{(N;k)}$, 
which are defined as the tensors of rank $k$ with components 
\begin{eqnarray}
\label{RCUMN}
\Sigma^{(N;k)}_{\alpha_1,\ldots,\alpha_k} =
N^{k-1} \mediac{\nu_{\alpha_1} \ldots \nu_{\alpha_k}},
\qquad \alpha_1, \ldots, \alpha_k \in \mathscr{H},
\end{eqnarray}
and let $\Sigma^{(k)}_{\alpha_1,\ldots,\alpha_k}$ be their asymptotic values  
\begin{eqnarray}\fl
\label{RCUM}
\Sigma^{(k)}_{\alpha_1,\ldots,\alpha_k} =
\lim_{N\to \infty}
N^{k-1} \mediac{\nu_{\alpha_1} \ldots \nu_{\alpha_k}}=
\lim_{N\to \infty}
\frac{1}{N} \mediac{N_{\alpha_1} \ldots N_{\alpha_k}}.
\end{eqnarray}
These limits exist and are finite since  
the irreducible and finite Markov chain formed by the evolving configurations 
has a finite correlation length $N_c$ with respect to the number of jumps $N$.
As a consequence, up to corrections exponentially small in $N/N_c$, 
for $N\to\infty$ we have 
\begin{eqnarray}
\label{LOGP1}
\log {\widetilde{\mathcal{P}}}_N(\bm{q}) &=
N \sum_{k=1}^{\infty} \frac{i^k}{k!} 
\sum_{\alpha_1\in\mathscr{H}} \ldots \sum_{\alpha_k\in\mathscr{H}}
N^{k-1}\mediac{\nu_{\alpha_1} \ldots \nu_{\alpha_k}}
q_{\alpha_1} \ldots q_{\alpha_k}
\nonumber \\
&= Ng(\bm{q}),
\end{eqnarray}
where the function $g(\bm{q})$ is independent of $N$.

By using the result (\ref{LOGP1}) and evaluating the integral 
in Eq. (\ref{FT}) by steepest descent, 
we find the asymptotic logarithm equality
\begin{equation}
P_N(N\bm{\nu}) \simeq \exp \left[ N \omega(\bm{\nu}) \right],
\end{equation}
where 
$\omega(\bm{\nu}) = g(\bm{q}^\mathrm{sp}) -i (\bm{q}^\mathrm{sp},\bm{\nu})$
is the function at the exponent evaluated in the saddle point
$\bm{q}^\mathrm{sp}(\bm{\nu})$. 
Equation (\ref{OMEGA}) follows immediately.

\section{The critical condition $\mathbf{W=1}$ in terms of the disorder}
\label{AppendiceB}
\setcounter{section}{2}
In this Appendix we show that the condition under which
the ground state of the system undergoes a quantum phase transition,
namely $W=1$, is equivalent to $WA=\mathcal{O}(1)$, where $A$
is the amount of disorder defined as $A=\overline{v}-v_\mathrm{min}$.
 
Let us consider a density $p(v)$ with $p(v_\mathrm{min})=0$ and 
suppose that all the statistical moments of $p(v)$ are finite, 
$\mediap{v^k}<\infty$, with $k$ non negative integer. 
Let us also suppose, for simplicity, that the density $p(v)$
is symmetric around the mean value $\overline{v}$ so that all 
the odd centered moments are zero and, furthermore, 
for any $v$ in the support of $p(v)$ we have
\begin{equation}
\label{cineq}
|\overline{v}-v|<|\overline{v}-v_\mathrm{min}|.
\end{equation} 
By using the geometric series, which is convergent due
to the above strict inequality, for any small positive $\delta$ we have
\begin{equation}\fl
\label{cint}
\int\frac{p(v)}{v-v_\mathrm{min}} dv =
\int_{v_\mathrm{min}}^{v_\mathrm{min}+\delta}
\frac{p(v)}{v-v_\mathrm{min}} dv + 
\frac{1}{\overline{v}-v_\mathrm{min}}
\int_{v_\mathrm{min}+\delta}^{v_\mathrm{max}}p(v)
\sum_{k=0}^{\infty}\frac{\left(v-\overline{v}\right)^{2k}}
{\left(\overline{v}-v_\mathrm{min}\right)^{2k}} dv.
\end{equation} 
In the interval $[v_\mathrm{min}+\delta,v_\mathrm{max}]$,
for sufficiently small positive $\delta$ we also have
\begin{equation}
\label{cineq2}
\frac{\overline{v}-v}{\overline{v}-v_\mathrm{min}}=
1-\frac{v-v_\mathrm{min}}{\overline{v}-v_\mathrm{min}}<
1-\frac{\delta}{\overline{v}-v_\mathrm{min}}<1.
\end{equation} 
Equation (\ref{cineq2}) implies that the series in the following inequality, 
\begin{equation}
\label{cint2}\fl
\sum_{k=0}^{\infty}
\int_{v_\mathrm{min}+\delta}^{v_\mathrm{max}}p(v)
\frac{\left(v-\overline{v}\right)^{2k}}
{\left(\overline{v}-v_\mathrm{min}\right)^{2k}} dv < 
\left(v_\mathrm{max}-v_\mathrm{min}-\delta\right)
\sum_{k=0}^{\infty}
\left(1-\frac{\delta}{\overline{v}-v_\mathrm{min}}\right)^{2k},
\end{equation} 
converges absolutely for any positive (sufficiently small) $\delta$, 
so that in the second term of the r.h.s. of
Eq. (\ref{cint}) we can exchange the order in which the operations
of integration and series appear.
Finally, by performing the limit $\delta\to 0$ and
using the fact that $p(v_\mathrm{min})=0$, we get 
\begin{equation}
\label{cW}
W=\frac{1}{\left(\overline{v}-v_\mathrm{min}\right)}
\sum_{k=0}^{\infty}
\frac{\mediap{(v-\overline{v})^{2k}}}{\left(\overline{v}-v_\mathrm{min}\right)^{2k}}.
\end{equation} 
From Eq.~(\ref{cW}) we see that $W$ is a decreasing function of
$A=\overline{v}-v_\mathrm{min}$ and $W=\mathcal{O}(1)/A$, so that
the critical condition amounts to $A=\mathcal{O}(1)$. 
By recalling that $v = V/|E_0^{(0)}|$, in terms of the non
rescaled potential the critical condition reads 
$\overline{V}-V_\mathrm{min}=\mathcal{O}(|{E_0^{(0)}}|)$.
This result is a generalization to many-body systems (in any dimension) 
of the result established for the single-particle Anderson models 
with dimension $d>1$ \cite{Anderson,FMSS}.

\end{document}